\documentclass[aps,prb,twocolumn,amsmath,amsfonts,amssymb,floatfix,letterpaper,showpacs]{revtex4}

\usepackage{graphicx}
\usepackage{ifpdf}

\usepackage[T1]{fontenc}

\newcommand{\ee}{\mathrm{e}}
\newcommand{\ii}{\mathrm{i}}
\newcommand{\dd}{\mathrm{d}}

\newcommand{\beq}[1]{\begin{equation}\label{#1}}
\newcommand{\eeq}{\end{equation}}
\newcommand{\refeq}[1]{Eq.~(\ref{#1})}

\newcommand{\refeqand}[2]{Eqs.~(\ref{#1}) and (\ref{#2})}

\newcommand{\beqm}[1]{\begin{multline}\label{#1}}

\newcommand{\punc}[1]{\,{\text{#1}}}
\newcommand{\sub}[1]{_{\text{#1}}}
\newcommand{\zero}{^{(0)}}

\newcommand{\Ham}{\mathcal{H}}
\newcommand{\J}{\mathcal{J}}

\newcommand{\Z}{\mathcal{Z}}

\newcommand{\del}{\boldsymbol{\nabla}}

\newcommand{\scAv}{\boldsymbol{\mathcal{A}}}
\newcommand{\scBv}{\boldsymbol{\mathcal{B}}}

\newcommand{\tr}{{\mathrm{T}}}

\newcommand{\nd}{^{\phantom{\dagger}}}
\newcommand{\ns}{^{\phantom{*}}}

\newcommand{\xh}{\hat{x}}
\newcommand{\yh}{\hat{y}}

\newcommand{\xhv}{\hat{\boldsymbol{x}}}
\newcommand{\yhv}{\hat{\boldsymbol{y}}}

\newcommand{\Tx}{\mathcal{T}_x}
\newcommand{\Ty}{\mathcal{T}_y}

\newcommand{\R}{\mathcal{R}}
\newcommand{\RR}{\mathbb{R}}
\newcommand{\Ix}{\mathcal{I}_x}
\newcommand{\Iy}{\mathcal{I}_y}
\newcommand{\Ixy}{\mathcal{I}_{xy}}
\newcommand{\IIx}{\mathbb{I}_x}
\newcommand{\IIy}{\mathbb{I}_y}
\renewcommand{\P}{\mathcal{P}}

\newcommand{\dsK}{\mathbb{K}}
\newcommand{\Km}{\mathbf{K}}

\renewcommand{\S}{\mathcal{S}}
\newcommand{\sm}{\mathbf{s}}
\renewcommand{\SS}{\mathbb{S}}
\newcommand{\Sigmam}{\boldsymbol{\Sigma}}

\newcommand{\kv}{\boldsymbol{k}}
\newcommand{\Kv}{\boldsymbol{K}}
\newcommand{\xv}{\boldsymbol{x}}
\newcommand{\rv}{\boldsymbol{r}}
\newcommand{\zerov}{\boldsymbol{0}}

\newcommand{\X}{\boldsymbol{X}}
\newcommand{\Y}{\boldsymbol{Y}}

\newcommand{\Hm}{\mathbf{H}}

\newcommand{\Mm}{\mathbf{M}}
\newcommand{\Mmzero}{\Mm_{\zerov}}
\newcommand{\Bm}{\mathbf{B}}

\newcommand{\Vv}{\mathbf{V}}
\newcommand{\Wv}{\mathbf{W}}

\newcommand{\Pv}{\mathbf{P}}
\newcommand{\Qv}{\mathbf{Q}}
\newcommand{\Av}{\mathbf{A}}

\newcommand{\Rv}{\boldsymbol{R}}
\newcommand{\TR}{\mathcal{T}_{\Rv}}

\newcommand{\alphav}{\boldsymbol{\alpha}}
\newcommand{\etam}{\boldsymbol{\eta}}
\newcommand{\Imatrix}{\boldsymbol{1}}
\newcommand{\gammam}{\boldsymbol{\gamma}}
\newcommand{\pim}{\boldsymbol{\pi}}

\newcommand{\BZ}{\mathfrak{B}}
\newcommand{\BL}{\BZ\sub{L}}
\newcommand{\BM}{\BZ\sub{M}}
\newcommand{\BN}{\BZ\sub{N}}

\newcommand{\at}{\tilde{a}}
\newcommand{\mut}{\nu}

\newcommand{\Psit}{\tilde{\Psi}}

\newcommand{\dkv}{\frac{\dd^2 \kv}{(2\pi)^2}}

\newcommand{\MSG}{\mathfrak{G}}
\newcommand{\MSGr}{\mathfrak{H}}

\newcommand{\pb}{\bar{p}}

\newcommand{\link}{\!\!\rightarrow\!\!}

\newcommand{\putinscaledfigure}[1]{\begin{center}\includegraphics[width=\columnwidth]{#1}\end{center}}

\usepackage{color}

\begin{document}

\title{Bogoliubov theory of interacting bosons on a lattice in a synthetic magnetic field}

\author{Stephen Powell}
\author{Ryan Barnett}
\author{Rajdeep Sensarma}
\author{Sankar Das Sarma}
\affiliation{Joint Quantum Institute and Condensed Matter Theory Center, Department of Physics, University of Maryland, College Park, MD 20742, USA}

\begin{abstract}
We consider theoretically the problem of an artificial gauge potential applied to a cold atomic system of interacting neutral bosons in a tight-binding optical lattice. Using the Bose-Hubbard model, we show that an effective magnetic field leads to superfluid phases with simultaneous spatial order, which we analyze using Bogoliubov theory. This gives a consistent expansion in terms of quantum and thermal fluctuations, in which the lowest order gives a Gross-Pitaevskii equation determining the condensate configuration. We apply an analysis based on the magnetic symmetry group to show how the spatial structure of this configuration depends on commensuration between the magnetic field and the lattice. Higher orders describe the quasiparticle excitations, whose spectrum combines the intricacy the Hofstadter butterfly with the characteristic features of the superfluid phase. We use the depletion of the condensate to determine the range of validity of our approximations and also to find an estimate for the onset of the Mott insulator phase. Our theory provides concrete experimental predictions, for both time-of-flight imagery and Bragg spectroscopy.
\end{abstract}


\maketitle

\section{Introduction}
\label{SecIntroduction}

One of the main goals of experiments with cold atoms has been to extend to new contexts physical effects that are familiar from the study of condensed matter. From this point of view, a particularly important area has been the study of vortices in superfluid systems, previously considered in the context of type-II superconductors \cite{Tinkham} and superfluid helium.\cite{Donnelly} When a trapped superfluid comprised of weakly interacting bosonic atoms is caused to rotate, it forms a lattice of vortices, a remarkably clear demonstration of the quantization of circulation.\cite{Donnelly,Cooper}

In a corotating reference frame, the (neutral) atoms experience a Coriolis force of the same form as the Lorentz force on charged particles in a magnetic field.\cite{Cooper} This magnetic analogy has inspired considerable experimental effort to achieve effective magnetic fields large enough to reach the quantum Hall regime, and corresponding theoretical work to extend the theory of the quantum Hall effect to the context of trapped bosons.\cite{Cooper}

Much recent effort has been directed towards combining effective magnetic fields and optical lattices, both to increase the stability of the experimental systems, and because of interesting physical effects expected in the presence of a lattice. The first experiments with rotating lattices used masks to produce parallel beams, whose subsequent interference formed the optical lattice potential. Mechanical rotation of the mask caused the interference pattern to rotate, and a density of vortices comparable to the density of lattice sites was achieved.\cite{Tung} Mechanical instabilities limited the lattice strength, however, restricting to the regime where the vortices are weakly pinned.\cite{Reijnders}

More recent experiments\cite{Williams} have replaced the mask with an acousto-optic modulator, allowing for considerably deeper lattices and lower temperatures. For a sufficiently deep lattice, a single-band approximation becomes valid, and the Bose-Hubbard model\cite{FWGF,Jaksch} gives a good description of the physics. Experiments with Raman lasers \cite{Spielman,Lin,LinNature} allow for effective magnetic fields without rotation, instead using coupling between internal atomic states to imprint the required geometric phases. A static optical lattice can be applied to such an arrangement, giving an effective magnetic field within the laboratory frame. Various other proposals have been made to induce phases directly within the lattice,\cite{Dum,JakschZoller,Mueller,Sorensen,Gerbier} some of which have the advantage of producing a perfectly commensurate flux density.

A major goal of these experiments is, as noted above, to reach the quantum Hall regime,\cite{Hafezi,Palmer,Moeller} in which the effective flux density (in units of the flux quantum $\phi_0 = 2\pi \hbar /Q$ where $Q$ is the effective particle charge) is comparable to particle density. Continuum systems in this regime exhibit a sequence of incompressible phases, the integer and fractional quantum Hall states, and, in the presence of a lattice, closely related physics has been predicted in certain areas of the phase diagram.\cite{Moeller}

Systems of bosons also support compressible superfluids, which are more closely related to the phases in the absence of a magnetic field, and are likely competitors with the quantum Hall states in the phase diagram. Detailed study of these phases is beneficial for the search for quantum Hall physics, both to calibrate experiments and to understand this competition, but their nontrivial properties and phenomena, especially in comparison with conventional superfluids, make them worthy of considerable interest in their own right.

These properties are inherited from the remarkable structure of the corresponding noninteracting problem, a single particle moving on a tight-binding lattice in the presence of a uniform magnetic field.\cite{Luttinger,Harper,Wannier,Zak,Hofstadter,Obermair} The density of states exhibits a fractal structure known as the `Hofstadter butterfly' (see Section~\ref{SecSpectrum}, and in particular Figure~\ref{FigButterfly}), with the spectrum depending sensitively on $\alpha$, the magnetic flux per plaquette of the lattice (measured in units of $\phi_0$). For rational $\alpha = p/q$ (with $p$ and $q$ coprime), there are $q$ bands, and each state is $q$-fold degenerate. This degeneracy can be understood in terms of the `magnetic symmetry group',\cite{Zak} which takes into account the modification of the symmetries inherent in making a particular choice for the gauge potential.

The present work introduces a theoretical approach that starts with the noninteracting Hofstadter spectrum and treats the superfluid using Bogoliubov theory. This gives a mean-field description of the superfluid phase of the Bose-Hubbard model, in which the interplay between the `magnetic' vorticity and the lattice potential (analogous to pinning effects for a weak lattice\cite{Reijnders,Goldbaum}) leads to real-space density modulations (`supersolidity'). The theory furthermore provides a consistent expansion in terms of both thermal and quantum fluctuations, and describes the quasiparticle spectrum above the condensate, which combines features of the Hofstadter butterfly and the Goldstone mode characteristic of superfluid order. We present detailed calculations within the superfluid phase, based on the microscopic Hamiltonian, which include predictions for both time-of-flight and Bragg spectroscopy measurements. A brief outline of our methods and results has been presented elsewhere.\cite{BogvLetter}

Bogoliubov theory has previously been applied to unfrustrated superfluids in optical lattices,\cite{Rey2} and also more recently to a proposed `staggered-flux' model,\cite{Lim} which shares some features of the present analysis. \DJ uri\'c and Lee\cite{Duric} have applied a spin-wave analysis to the system considered here, using a real-space perspective that provides results consistent with ours. A related analysis has been applied to fermions,\cite{Zhai} which similarly show spatial order in the paired superfluid phase.

The real-space structure of the condensate also bears many similarities to other systems with phase coherence in a magnetic field. In particular, superconducting lattices with applied fields support states with current patterns similar to those discussed in Section~\ref{SecMFTheory} below,\cite{Alexander} and experiments on these systems have clearly demonstrated the effect of the Hofstadter spectrum on the superconducting transition.\cite{Pannetier} Frustrated Josephson junction arrays,\cite{Teitel,Halsey} where the charging energy can become comparable to the Cooper-pair tunneling amplitude, provide a close analogue of the present system, and implementations using cold atoms have been proposed.\cite{Polini,Kasamatsu}

Previous work on this system has also considered the Mott insulator that should exist for strong interactions and commensurate density.\cite{FWGF,Jaksch,Oktel,Umucalilar,Goldbaum} This phase, which is very similar to its analogue in the absence of a magnetic field, is favored when the Hubbard $U$ interaction suppresses number fluctuations and eliminates the coherence between neighboring sites of lattice. The transition between this phase and the superfluid has been studied using the Gutzwiller ansatz\cite{Scarola,Oktel,Umucalilar,Goldbaum,Goldbaum2,Lundh} and effective field-theory methods,\cite{Sinha} with the main effect being an enhancement of the insulator phase due to the frustration\cite{Moessner} of the hopping.

In this work, we restrict consideration to uniform magnetic fields. By modulating the effective field in real space, it is possible to produce states with nontrivial topological properties.\cite{Stanescu} These include topological insulators and metals, which display similar physics to the single-particle states underlying the phenomena described here. Recent work has also addressed systems of interacting bosons in spatially varying magnetic fields.\cite{Saha}

Although the framework that we present has broad applicability, our specific model includes several simplifications. First, we use a single-band Hubbard model, incorporating only the lowest band of the optical-lattice potential. (One effect of the magnetic field is to split the tight-binding spectrum into multiple `Hofstadter bands', and these are included exactly in our approach.) While this is certainly not a valid approximation for the shallow lattices of earlier experiments,\cite{Tung} the single-band limit is approached by more recent work with rotating lattices,\cite{Williams} and should be readily achievable in combination with Raman-induced gauge potentials.\cite{Lin,LinNature}

We also assume a spatially uniform system with exactly rational $\alpha$. The main effect of the finite trap size is to wash out the small-scale fractal structure of the Hofstadter butterfly,\cite{Hofstadter,JakschZoller} so we expect most of our results to be valid for any $\alpha$. The experimental consequences of nonuniformity and incommensurate magnetic flux will be addressed in more detail in future work.

\subsection{Outline}
\label{SecOutline}

Our approach is as follows: In Section~\ref{SecSingleParticle}, we use a symmetry analysis to find the full spectrum of single-particle states, described by bosonic annihilation operators $a_{\kv\ell\gamma}$, defined below in \refeq{bfromakl}. We then apply the Bogoliubov ansatz,\cite{Bogoliubov,AGD}
\beq{BogvAnsatz}
a_{\kv \ell \gamma} = A_{\ell \gamma} (2\pi)^2 \delta^2(\kv) + \at_{\kv \ell \gamma}\punc{,}
\eeq
where $A_{\ell\gamma}$ is a set of c-numbers giving the condensate order parameter, and the operators $\at_{\kv\ell\gamma}$ describe the residual bosons outside the condensate. An expansion in terms of these operators is then developed, which in physical terms is an expansion in fluctuations around mean-field theory. In Section~\ref{SecMFTheory}, we address the lowest-order term, which involves only the constants $A_{\ell\gamma}$ and leads to a time-independent Gross-Pitaevskii equation for the condensate configuration.

Higher-order terms, describing the spectrum of low-energy excitations and their interactions, are treated in Section~\ref{SecBogv}. In this work, we truncate the expansion at quadratic order, leading to a theory of noninteracting Bogoliubov quasiparticles. For this approximation to be reasonable, one requires a sufficiently low density of quasiparticles, and we determine the regime of validity by calculating the condensate depletion.

In Section~\ref{SecExperiment}, we describe the consequences for experiments, giving predictions for both time-of-flight imagery and Bragg spectroscopy. We conclude with discussion in Section~\ref{SecDiscussion}. Some details of the Bogoliubov transformation and the symmetry analysis of the interacting spectra are given in Appendices~\ref{AppBogoliubov} and \ref{AppSymmetries}. In Appendix~\ref{AppAnalytics} we provide more details and analytic results for the simplest case, $\alpha = \frac{1}{2}$.

\subsection{Hamiltonian}
\label{SecHamiltonian}

The single-band Bose-Hubbard model in the presence of a static magnetic field can be written\cite{JakschZoller} as
\beq{Hamiltonian}
\Ham = -t \sum_{\langle i j \rangle}\left(\ee^{\ii \Phi_{ij}} b_j^\dagger b_i\nd + \text{h.c.}\right) + \frac{U}{2}\sum_j n_j(n_j - 1)\punc{,}
\eeq
where $b_j$ and $n_j = b^\dagger_j b\nd_j$ are the annihilation and number operators respectively for site $j$. We treat a two-dimensional square lattice, and assume hopping with amplitude $t$ only between nearest neighbors, denoted $\langle i j \rangle$. The second term gives an on-site interaction with strength $U$; the approach described here can straightforwardly be extended to incorporate more complicated interactions, including between particles on different sites. (Note that we treat electrically neutral particles in the presence of a synthetic magnetic field, and we will assume that there are no long-range interactions.)

The phase $\Phi_{ij}$ on the directed link from site $i$ to site $j$, to be denoted $i \link j$, can be expressed\cite{Luttinger} in terms of the magnetic vector potential $\scAv$ as
\beq{HoppingPhases}
\Phi_{ij} = Q \int_{\xv_i}^{\xv_j} \dd\rv \cdot \scAv(\rv)\punc{.}
\eeq
The integral is to be taken along the straight-line path between the positions $\xv_i$ and $\xv_j$ of the two sites; one has $\Phi_{ij} = -\Phi_{ji}$. While the individual phases depend on the choice of gauge, the lattice curl
\beq{MagneticFlux}
\sum_{ij \circlearrowleft \square} \Phi_{ij} = Q \int_\square \dd^2 \rv \cdot \del \times \scAv(\rv)
\eeq
is gauge-independent, where the sum is over links enclosing a given plaquette $\square$ in a counterclockwise sense.

The integral in \refeq{MagneticFlux} is simply the magnetic flux through the plaquette, given by $|\scBv|a^2$, assuming a uniform magnetic field $\scBv = \del \times \scAv$ perpendicular to a square lattice with spacing $a$. The dimensionless flux per plaquette is defined by $\alpha = |\scBv|a^2 / \phi_0$, and completely specifies the effect of the magnetic field. We will henceforth use units where $\hbar = a = 1$, and take the effective charge on the bosons as $Q = +1$, so the flux quantum is simply $\phi_0 = 2\pi$, and \refeq{MagneticFlux} becomes
\beq{MagFlux2}
\sum_{ij \circlearrowleft \square} \Phi_{ij} = 2\pi \alpha\punc{.}
\eeq

\begin{figure}
\putinscaledfigure{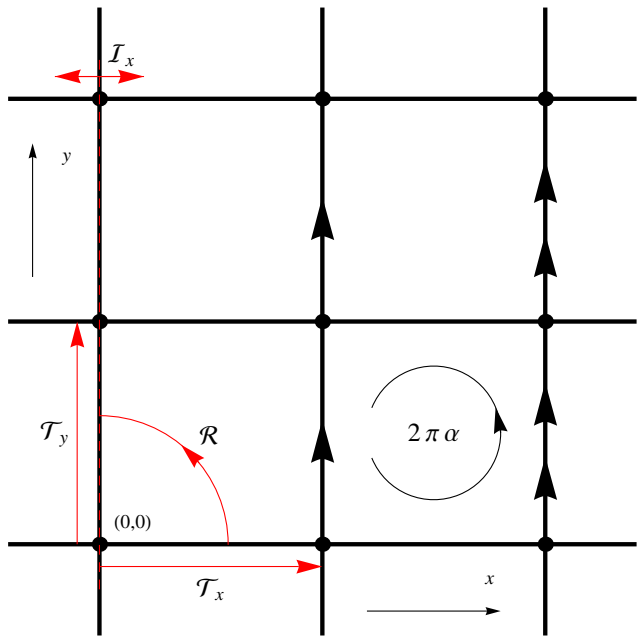}
\caption{\label{FigLandauGaugeSymmetries}An illustration of the vector potential $\Phi_{ij}$ in the Landau gauge, and the symmetries defined in Section~\ref{SecMSG}. The number of arrowheads on each link of the square lattice gives the value of $\Phi_{ij}$ on the directed link $i \link j$, in units of $2\pi\alpha$. The configuration is one possible choice obeying \refeq{MagFlux2}: going around any square plaquette in a counterclockwise sense, the number of forward arrows minus backward arrows is $+1$. In this choice of gauge, $\Phi_{ij}$ vanishes on all links in the $x$ direction. Any choice of gauge reduces the full symmetry of the square lattice, and the translation and rotation operations shown are only symmetries when accompanied by phase factors (see Section~\ref{SecMSG}). Reflections must be combined with time-reversal operations to preserve the direction of the applied magnetic field.}
\end{figure}

It is convenient theoretically, and also for recent experiments with effective gauge potentials,\cite{Lin,LinNature} to use the Landau gauge, in which the vector potential is parallel to one of the square lattice axes. We take the vector potential along $\yhv$, which is conventional in the theoretical literature, but it should noted that in the recent experiments of Lin et al.,\cite{Lin,LinNature} it is instead aligned with $\xhv$. Our choice of gauge is illustrated in Figure~\ref{FigLandauGaugeSymmetries}.

With this choice, one has $\ee^{\ii \Phi_{j,j+\xh}} = 1$ and $\ee^{\ii \Phi_{j,j+\yh}} = \ee^{2\pi \ii \alpha x_j}$, where (for example) $j+\xh$ denotes the site adjacent to $j$ in the $+\xhv$ direction. The kinetic term $\Ham_t$ then becomes
\begin{multline}
\label{HamtLandau}
\Ham_t = -t \sum_j\big[ b^\dagger_{j+\xh} b\nd_j + b^\dagger_{j-\xh} b\nd_j + \omega^{x_j} b^\dagger_{j+\yh} b\nd_j\\
{}+ \omega^{-x_j} b^\dagger_{j-\yh} b\nd_j \big]\punc{,}
\end{multline}
where $\omega = \ee^{2\pi \ii \alpha}$. Note that the phases act to `frustrate' the hopping, so that for noninteger $\alpha$ it is not possible to minimize the kinetic energy on every link of the lattice simultaneously.\cite{Moessner}

The transformation to an alternative gauge is implemented by applying a spatially-varying phase rotation to $b_j$. For example, in the symmetric gauge, which is more natural for a rotating lattice, one defines operators $\tilde{b}_j = \omega^{-x_j y_j/2}b_j$, so that the gauge field becomes $\ee^{\ii \Phi_{j,j+\xh}} = \omega^{-y_j/2}$ and $\ee^{\ii \Phi_{j,j+\yh}} = \omega^{x_j/2}$. The number operator $n_j = b^\dagger_j b\nd_j$ is invariant under any such gauge transformation, as required for a physically observable quantity, and so the interaction energy $\Ham_U$ is also invariant.

As noted above, we assume precisely rational $\alpha = p/q$ (with $p$ and $q$ coprime), so that $\omega^q = 1$, and the unit cell of $\Ham$ is $q \times 1$ sites. (In the symmetric gauge, the unit cell is considerably larger, containing $2q \times 2q$ sites.)

\section{Single-particle spectrum}
\label{SecSingleParticle}

We begin by describing in detail the spectrum of the single-particle kinetic term $\Ham_t$, given in \refeq{HamtLandau}. While these results\cite{Harper,Zak,Hofstadter} are well established (and have been for many decades), we will present them here in some detail, in order to introduce the concepts and formalism that will be central to our subsequent analysis.

The general structure of the spectrum for any $\alpha = p/q$ can be determined by considering the ``magnetic symmetry group'' (MSG; also known as the ``projective symmetry group'').\cite{Zak,Wen} This arises because any choice of gauge necessarily reduces the physical symmetry of the lattice and leads to a Hamiltonian that does not commute with the standard spatial symmetry operators. For example, while the (uniform) magnetic field is invariant under translation by a single lattice site in any direction, $\Ham_t$ is manifestly asymmetric under translations in the $x$ direction.

One can nonetheless define a group of operators in one-to-one correspondence with the physical symmetries, that obey the multiplication table apart from phase factors, and that commute with $\Ham_t$ (and indeed the full Hamiltonian $\Ham$). We define this group by specifying operators corresponding to the elementary translation, rotation, and reflection transformations from which the full group can be constructed.

\subsection{Magnetic symmetry group}
\label{SecMSG}

We first define the elementary translation operators $\Tx$ and $\Ty$, illustrated in Figure~\ref{FigLandauGaugeSymmetries}, in terms of their commutation with the annihilation operator $b_j$. With our choice of the Landau gauge, the Hamiltonian is symmetric under translations in the $y$ direction, and so one can define $\Ty$ by
\beq{Ty}
\Ty b_j = b_{j+\yh} \Ty\punc{.}
\eeq
In contrast, the $x$-dependent phase factors in \refeq{HamtLandau} imply that a pure translation does not commute with $\Ham_t$, and we instead define $\Tx$ by
\beq{Tx}
\Tx b_j = b_{j+\xh} \Tx \omega^{-y_j}\punc{.}
\eeq
The gauge field has periodicity $q$ in the $x$ direction, so the combination $\Tx^q$ obeys $\Tx^q b_j = b_{j+q\xh} \Tx^q$.

The multiplication relations of the MSG are equal up to phase factors to those of the ordinary spatial group. With total particle number $N = \sum_j n_j$, one finds
\beq{TxTy}
\Tx \Ty = \Ty \Tx \omega^N\punc{,}
\eeq
by induction, starting from the (totally symmetric) vacuum state. While the phase factors associated with individual operators are dependent on the choice of gauge, this relation is gauge independent.

Besides translations, it is useful to consider rotation and reflection operations. We define the unitary operator $\R$ giving a rotation by $90^\circ$ counterclockwise about the site at the origin (see Figure~\ref{FigLandauGaugeSymmetries}):
\beq{R}
\R b_j = b_{\RR j} \omega^{x_j y_j} \R\punc{,}
\eeq
where $j \rightarrow \RR j$ under the rotation ($x_{\RR j} = -y_j$, $y_{\RR j} = x_j$). The phase factor again ensures that $[\R,\Ham_t] = 0$.

For reflection operators, the situation is somewhat different, since the magnetic field explicitly breaks chirality and time-reversal symmetry. While reflection reverses chirality and so is not a symmetry of the Hamiltonian, a combination of reflection and time reversal restores the appropriate sense of circulation and remains a symmetry in the presence of the field. One can therefore define antiunitary operators corresponding to such combinations; we define $\Ix$ and $\Iy$ for the transformations obeying $x_{\IIx j} = -x_j$ and $y_{\IIy j} = -y_j$ respectively. No phase factors are required in these cases.

It is also useful to define the combinations $\P = \R^2 = \Ix\Iy$ and $\Ixy = \R \Iy$. The former gives inversion about the origin, $\xv \rightarrow -\xv$, which in two dimensions is a proper rotation and hence represented by a unitary operator. The antiunitary operator $\Ixy$ gives reflection in the line $y = x$.

Note that the interaction term $\Ham_U$ is a function only of the gauge-invariant combination $n_j = b^\dagger_j b\nd_j$ and so is unaffected by the phase factors included in expressions such as \refeq{Tx}. Any interaction term with the full symmetry of the lattice, such as the explicit example in \refeq{Hamiltonian}, is therefore invariant under the magnetic symmetry group.

\subsection{Momentum-space operators}
\label{SecMomentumSpace}

While the unit cell of the Hamiltonian $\Ham_t$ contains $q$ sites and hence gives a reduced Brillouin zone, it is useful to construct momentum space operators based on the full lattice Brillouin zone $\BL$: $-\pi \le k_x,k_y < \pi$. The momentum-space annihilation operator $b_{\kv}$ is defined by
\beq{bFT}
b_{\kv} = \sum_j \ee^{-\ii \xv_j \cdot \kv} b_j\punc{,}
\eeq
so that the commutator is given by
\beq{kComm}
[b\nd_{\kv}, b^\dagger_{\kv'}] = (2\pi)^2 \delta^2([\kv - \kv']_{\BL})\punc{.}
\eeq
Here and throughout, we use the notation $[\kv]_{\BZ}$ to denote the momentum $\kv$ reduced to the Brillouin zone $\BZ$ by the addition of an appropriate lattice vector. We also use the shorthand notation $[x]_q = x \bmod q$.

Written in momentum space, the Hamiltonian $\Ham_t$ is
\begin{widetext}
\beq{Hamtk}
\Ham_t = -t \int_{\kv \in \BL}\! \dkv \bigg( 2 \cos k_x \: b^\dagger_{\kv} b\nd_{\kv} \\
{}+ \ee^{-\ii k_y} b^\dagger_{[\kv + \X]_{\BL}} b\nd_{\kv} + \ee^{\ii k_y} b^\dagger_{[\kv - \X]_{\BL}} b\nd_{\kv} \bigg)\punc{,}
\eeq
where $\X = 2\pi \alpha \xhv$. The enlarged unit cell ($q \times 1$ sites) of the Hamiltonian allows mixing between momentum states that coincide when reduced to the magnetic Brillouin zone $\BM$: $-\pi \le k_y < \pi$, $-\pi/q \le k_x < \pi/q$.

The operators $b_{\kv}$ commute with $\Ty$,
\beq{Tybk}
\Ty b_{\kv} = \ee^{\ii k_y} b_{\kv} \Ty\punc{,}
\eeq
but the phase factor in the definition of $\Tx$ causes it to mix momenta,
\beq{Txbk}
\Tx b_{\kv} = \ee^{\ii k_x} b_{[\kv + \Y]_{\BL}} \Tx\punc{,}
\eeq
where $\Y = 2\pi \alpha \yhv$. Since $\Tx$ commutes with $\Ham_t$, this implies degeneracies in the single-particle spectrum between points separated by $\Y$. To make this transparent, it is convenient to define the doubly reduced Brillouin zone $\BN$: $-\pi/q \le k_x,k_y < \pi/q$.

Any point within $\BL$ can be specified as $[\kv + \ell \X + n \Y]_{\BL}$, where $\kv \in \BN$ and $n$ and $\ell$ are integers from $0$ to $q-1$, so $\Ham_t$ can be rewritten as
\beq{Hamtk2}
\Ham_t = \int_{\kv \in \BN} \! \dkv \sum_{\ell=0}^{q-1} \sum_{n,n'=0}^{q-1} b^\dagger_{[\kv + n \X + \ell \Y]_{\BL}} H_{nn'}([\kv + \ell \Y]_{\BL}) b\nd_{[\kv + n' \X + \ell \Y]_{\BL}}\punc{,}
\eeq
\end{widetext}
where the $q \times q$ matrix $\Hm(\kv)$ has elements (for $q>2$)
\beq{Hmatrix}
H_{nn'}(\kv) = -t\times\begin{cases}
\ee^{+\ii k_x}\omega^n + \ee^{-\ii k_x}\omega^{-n} & \text{if $n' = n$}\\
\ee^{+\ii k_y} & \text{if $n' = [n+1]_q$}\\
\ee^{-\ii k_y} & \text{if $n' = [n-1]_q$}\\
0 & \text{otherwise.}
\end{cases}
\eeq
(If $q = 2$, $[n+1]_q = [n-1]_q$ and $H_{01} = H_{10} = -2t\cos k_y$.)

Noting that $H_{nn'}([\kv + \ell \Y]_{\BL}) = \omega^{(n-n')\ell}H_{nn'}(\kv)$, we diagonalize $\Ham_t$ by writing
\beq{bfromakl}
b_{[\kv + n \X + \ell \Y]_{\BL}} = \omega^{-n\ell} \sum_{\gamma} \psi_{\gamma n}(\kv) a_{\kv\ell\gamma}\punc{,}
\eeq
where $a_{\kv\ell\gamma}$ is the annihilation operator for a single-particle state labeled by band index $\gamma \in \{1, \ldots, q\}$. For each $\kv \in \BN$, $\psi_{\gamma n}(\kv)$ is an eigenvector of $\Hm(\kv)$ with eigenvalue $\epsilon_\gamma(\kv)$. The $q$ bands can be understood from the `folding' of the Brillouin zone due to the reduced translation symmetry of $\Ham$.

It should be noted that the annihilation operator $a_{\kv\ell\gamma}$ has momentum $\kv$ when referred to $\BN$, but momentum $\kv + \ell\Y$ in the magnetic Brillouin zone $\BM$. For each $\kv \in \BN$, the eigenvectors $\psi_{\gamma n}(\kv)$ corresponding to different bands are orthogonal, so the operators $a_{\kv\ell\gamma}$ obey canonical commutation relations,
\beq{aComm}
[a\nd_{\kv\ell\gamma} a^\dagger_{\kv'\ell'\gamma'}] = (2\pi)^2\delta^2([\kv - \kv']_{\BN})\delta_{\ell\ell'}\delta_{\gamma\gamma'}\punc{.}
\eeq

The single-particle Hamiltonian can finally be rewritten as
\beq{Hamta}
\Ham_t = \int_{\kv \in \BN} \! \dkv \sum_{\ell=0}^{q-1}\sum_{\gamma} \epsilon_\gamma(\kv) a^\dagger_{\kv\ell\gamma} a\nd_{\kv\ell\gamma} \punc{.}
\eeq
The single-particle energy $\epsilon_\gamma(\kv)$ is independent of $\ell$, so every state is (at least) $q$-fold degenerate. The band labels $\gamma$ can be arranged so that $\epsilon_\gamma(\kv) \le \epsilon_{\gamma + 1}(\kv)$ for every $\kv$ and $\epsilon_\gamma(\kv)$ is a continuous function of $\kv$.

Using \refeqand{Tybk}{Txbk}, one finds the effect of the operators $\Ty$ and $\Tx$ as
\begin{align}
\label{Tyakl}
\Ty a_{\kv\ell\gamma} &= \ee^{\ii k_y} \omega^\ell a_{\kv\ell\gamma} \Ty\\
\label{Txakl}
\Tx a_{\kv\ell\gamma} &= \ee^{\ii k_x} a_{\kv[\ell+1]_q\gamma} \Tx\punc{.}
\end{align}
The operator $\Tx$ therefore transforms one degenerate single-particle state into another, and it is this symmetry that enforces the degeneracy.

Determining the effects of the rotation and reflection operators $\R$, $\Ix$, and $\Iy$ is somewhat more involved.\cite{Balents} Considering first $\R$, its commutation with $\Ham_t$ implies that $\R a_{\kv \ell \gamma} \R^\dagger$ can be written in terms of $a_{(\RR \kv) \ell' \gamma}$ in the same band $\gamma$. With an appropriate choice of the arbitrary phase of the eigenvectors $\psi_{\gamma n}$ at the points $\kv$ and $\RR \kv$, one finds
\beq{Rak}
\R a_{\kv \ell \gamma} = \frac{1}{\sqrt{q}} \ee^{-\ii \phi_\gamma} \sum_{\ell'} \omega^{\ell \ell'} a_{(\RR \kv) \ell' \gamma} \R\punc{,}
\eeq
where the phase $\phi_\gamma$ is independent of $\kv$. The requirement that $\R^4 = 1$ implies that $\phi_\gamma$ is a multiple of $\pi/2$ for all bands $\gamma$. One similarly finds
\begin{align}
\label{Ixak}
\Ix a_{\kv \ell \gamma} &= a_{(\IIy \kv)[-\ell]_q \gamma} \Ix\\
\label{Iyak}
\Iy a_{\kv \ell \gamma} &= \ee^{-2\ii \phi_\gamma} a_{(\IIx \kv)\ell\gamma} \Iy\\
\label{Pak}
\P a_{\kv \ell \gamma} &= \ee^{-2\ii\phi_\gamma}a_{(-\kv)[-\ell]_q\gamma}\P
\punc{.}
\end{align}
Note that $\ee^{-2\ii\phi_\gamma} = \pm 1$ is real. Similar expressions, such as
\beq{psiRk}
\psi_{\gamma n}(\RR\kv) = \frac{1}{\sqrt{q}} \ee^{-\ii\phi_\gamma}\sum_{n'}\omega^{nn'}\psi_{\gamma n'}(\kv)\punc{,}
\eeq
relate the eigenvectors $\psi_{\gamma n}(\kv)$ at symmetry-equivalent momenta $\kv$.

Applying these operators to the Hamiltonian in the form of \refeq{Hamta} immediately shows that the single-particle dispersion $\epsilon_\gamma(\kv)$ is symmetric under the corresponding transformations of the momentum $\kv$. For example, $\epsilon_\gamma(\RR \kv) = \epsilon_\gamma(\kv)$, which implies that the dispersion has the full four-fold rotation symmetry of the lattice, despite the reduced symmetry of $\Ham_t$.

\subsection{Spectrum}
\label{SecSpectrum}

\begin{figure}
\putinscaledfigure{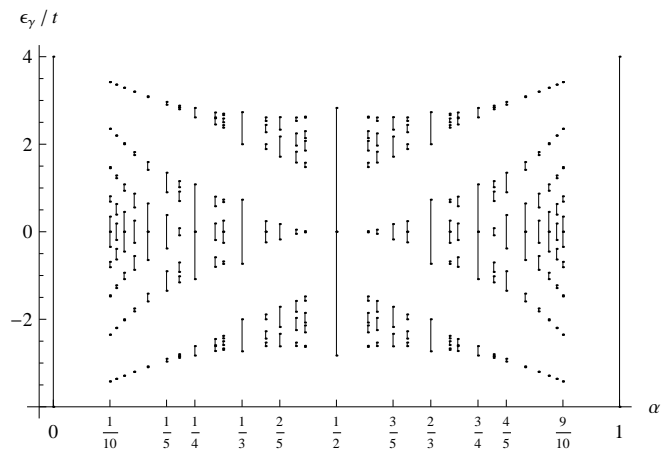}
\caption{\label{FigButterfly}The Hofstadter butterfly,\cite{Hofstadter} a plot of the single-particle energies $\epsilon_\gamma$ as a function of flux $\alpha$. The butterfly has a fractal structure, but only a finite set of $\alpha = p / q$ can be plotted; for clarity, we restrict to $q \le 10$. Points mark the top and bottom of each of the $q$ bands, which become increasingly narrow as $q$ becomes larger. In the limit $q\rightarrow \infty$ with $p$ fixed, or equivalently $\alpha \ll 1$, the low-lying bands become the Landau levels of the continuum.}
\end{figure}

In summary, for $\alpha = p/q$, the single-particle spectrum consists of $q$ bands, labeled by $\gamma$, with each state $q$-fold degenerate. These degenerate states have energy $\epsilon_\gamma(\kv)$ and momentum $[\kv + \ell \Y]_{\BM}$, with $\ell \in \{ 0,\ldots,q-1\}$. Figure~\ref{FigButterfly} shows the `Hofstadter butterfly', \cite{Hofstadter} a plot of the allowed single-particle energies $\epsilon_\gamma$ (for any momentum $\kv$) as a function of the flux $\alpha$. The plot has a fractal structure \cite{Hofstadter} that is sensitively dependent on $\alpha$, and for clarity only rational $\alpha = p / q$ with $q \le 10$ have been included. For each $\alpha = p/q$, points mark the top and bottom of each of the $q$ bands.

As can be seen in Figure~\ref{FigButterfly}, most of the bands are separated by nonzero gaps. The only exceptions are the two central bands for $q$ even, which touch exactly at the point of zero energy. These occur at $\kv = \zerov$ for $q$ an integer multiple of $4$, and at the corner of $\BN$, $\kv_{\mathrm{X}} = \frac{\pi}{q}\xhv + \frac{\pi}{q}\yhv$, otherwise. For both, the spectrum has a linear `Dirac-cone' dispersion near the degeneracy point.\cite{Kohmoto}

\begin{figure}
\putinscaledfigure{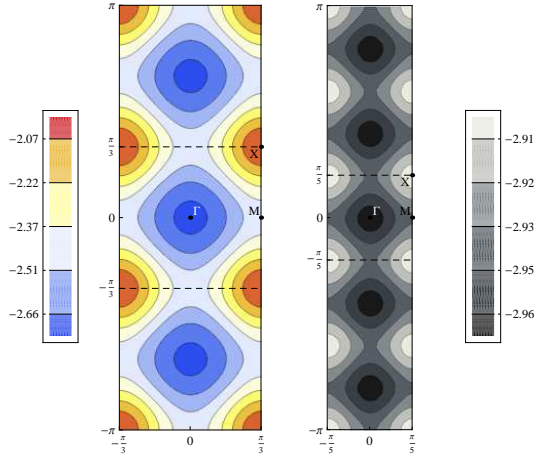}
\caption{\label{FigContours}Contour plot of the lowest band of the noninteracting single-particle spectrum $\frac{1}{t}\epsilon_1(\kv)$, for $\alpha = \frac{p}{q} = \frac{1}{3}$ (left) and $\frac{1}{5}$ (right). In both cases, the spectrum is plotted in the magnetic Brillouin zone $\BM$, in which $-\frac{\pi}{q} \le k_x < \frac{\pi}{q}$ and $-\pi \le k_y < \pi$. Points separated by momentum $\Y = 2\pi\alpha\yhv$ are degenerate; identifying these gives the doubly reduced Brillouin zone $\BN$, $-\frac{\pi}{q}\le k_x,k_y < \frac{\pi}{q}$, indicated by the horizontal dashed lines. Note that the lowest band is considerably narrower in the case $\alpha = \frac{1}{5}$, as is also evident in Figure~\ref{FigButterfly}.}
\end{figure}

In all other cases, including the lowest band for all $\alpha$, the dispersion is quadratic near its minimum. The minimum of the lowest band always occurs at $\kv = \zerov$, as can be shown using the Perron-Frobenius theorem, and the effective mass near this point can be found by perturbation theory for small $\kv$. In all cases, the coefficients are equal in the $x$ and $y$ directions (i.e., the effective mass tensor is proportional to the unit matrix), a straightforward consequence of the symmetry $\R$. Figure~\ref{FigContours} shows the lowest band of the noninteracting dispersion, $\epsilon_1(\kv)$, for $\alpha = \frac{1}{3}$ and $\frac{1}{5}$. In both cases, the dispersion is quadratic and isotropic near the top and bottom of the band.

\subsection{Interactions}
\label{SecInteractions}

The operators $a_{\kv\ell\gamma}$ defined in Section~\ref{SecMomentumSpace} are chosen to diagonalize the kinetic energy operator $\Ham_t$. To incorporate the effects of the interaction term $\Ham_U$, this must also be expressed in terms of these operators. This involves the straightforward process of substituting \refeqand{bFT}{bfromakl} into $\Ham_U$, and can be performed for any choice of interaction. Our explicit calculations are for the on-site Hubbard interaction in \refeq{Hamiltonian}, appropriate to bosons in a deep optical lattice.

A general quartic interaction can be written in terms of the operators $a_{\kv\ell\gamma}$ as
\beq{HamUa}
\Ham_U = \int_{\kv_1\cdots\kv_4}\sum_{\ell_1\cdots\ell_4} \sum_{\gamma_1\cdots\gamma_4} u\, a^\dagger_{\kv_1\ell_1\gamma_1} a^\dagger_{\kv_2\ell_2\gamma_2}a\nd_{\kv_3\ell_3\gamma_3} a\nd_{\kv_4\ell_4\gamma_4}\punc{,}
\eeq
where the coefficient $u$ is a function of the four sets of indices $\kv$, $\ell$, and $\gamma$. It can be chosen symmetric under exchange of the first two ($1 \leftrightarrow 2$) or last two sets ($3 \leftrightarrow 4$), and the requirement that $\Ham_U$ be hermitian implies that $u(3,4,1,2) = u^*(1,2,3,4)$.

A translation-invariant interaction conserves momentum, so the coupling coefficient $u$ is nonzero only if
\beq{ConsOfMom}
\big[(\kv_1+\kv_2-\kv_3-\kv_4) + (\ell_1+\ell_2-\ell_3-\ell_4)\Y \big]_{\BM} = \zerov\punc{.}
\eeq
Note that this allows for umklapp processes where the net momentum is zero only when reduced to $\BN$. Factoring out $(2\pi)^2\delta^2([\kv_1+\kv_2-\kv_3-\kv_4]_{\BN})$ gives $\bar{u}$:
\begin{widetext}
\beq{ubar}
\bar{u}(\{\kv\},\{\ell\},\{\gamma\}) = \frac{U}{2}\delta_{\{\ell\}}\sum_{n_1 \cdots n_4} \delta_{\{n\}} \psi^*_{\gamma_1 n_1}(\kv_1)\psi^*_{\gamma_2 n_2}(\kv_2) \psi\ns_{\gamma_3 n_3}(\kv_3)\psi\ns_{\gamma_4 n_4}(\kv_4) \omega^{n_1 \ell_1 + n_2\ell_2 - n_3\ell_3 - n_4 \ell_4}\punc{,}
\eeq
where $\delta_{\{\ell\}}$ and $\delta_{\{n\}}$ denote Kronecker deltas enforcing \refeq{ConsOfMom} and a similar constraint on $n_{1\cdots4}$. The complicated structure of the single-particle states gives $\bar{u}$ a nontrivial dependence on the momenta $\kv$ of the interacting particles, despite the choice of a purely on-site interaction.

Note that $\bar{u}$ is, up to a phase factor that is only nontrivial for umklapp processes, only dependent on two of the $\ell$'s:
\beq{ubar2}
\bar{u}(\ell_1,\ell_2,\ell_3,\ell_4) = \ee^{-\ii p \pb \ell_1(\kv_1 + \kv_2 - \kv_3 - \kv_4).\xhv} \bar{u}(0,[\ell_2-\ell_1]_q, [\ell_3-\ell_1]_q,[\ell_4 - \ell_1]_q)\punc{,}
\eeq
where $\pb$ is the modulo-$q$ reciprocal of $p$, the integer such that $[p \pb]_q = 1$ and $0 < \pb < q$. This identity, and the momentum-conservation constraint of \refeq{ConsOfMom} are consequences of the symmetry of the interactions under $\Tx$ and $\Ty$. Further constraints on the coefficients $u$ result from the symmetry properties of the eigenvectors $\psi_{\gamma n}(\kv)$ under rotations and reflections. For example, requiring that $\P$ commutes with $\Ham_U$ and using \refeq{Pak} gives
\beq{uParitySymmetry}
\ee^{-2\ii \sum_{i=1}^4\phi_{\gamma_i}} u(\kv_1\ldots\kv_4,\ell_1\ldots\ell_4,\gamma_1\ldots\gamma_4) = \\ u(-\kv_1\ldots-\kv_4,[-\ell_1]_q\ldots[-\ell_4]_q,\gamma_1\ldots\gamma_4)
\punc{.}
\eeq
\end{widetext}
This implies that $u$ is odd in momentum, and hence vanishes for $\kv_{1\cdots 4} = \zerov$, for certain combinations of $\ell_{1\cdots 4}$ and $\gamma_{1\cdots 4}$. A detailed discussion of the restrictions imposed by symmetries has been presented by Balents et al.\cite{Balents}\ in a different context.

\section{Mean-field theory}
\label{SecMFTheory}

Having described the spectrum of noninteracting particles, we now turn to the effects of interactions on the many-body physics. As described in Section~\ref{SecOutline}, our approach will be based on the Bogoliubov theory, using the ansatz \refeq{BogvAnsatz} and performing an expansion in powers of the fluctuation operators. The Bogoliubov ansatz can be viewed as a statement about correlation functions in the superfluid phase, with the first term in \refeq{BogvAnsatz} giving the one-point correlation function, the condensate order parameter $\langle a_{\kv \ell \gamma}\rangle = A_{\ell\gamma}(2\pi)^2\delta^2(\kv)$. The zeroth-order term in the Bogoliubov expansion is given by neglecting higher-order connected correlation functions.

This mean-field theory can be viewed as a special case of that derived by using the Gutzwiller ansatz, which assumes a state $\prod_j |\psi_j\rangle$ that is factorizable in real space.\cite{Scarola,Oktel,Goldbaum,Umucalilar} In general, one allows $|\psi_j\rangle$ to be an arbitrary state within the on-site manifold, but our ansatz assumes a bosonic coherent state and is appropriate only within the superfluid phase.

Substituting the mean-field ansatz into the Hamiltonian, written as in \refeqand{Hamta}{HamUa}, gives the energy density
\begin{multline}
\label{MFT}
h_0 = \sum_{\ell,\gamma} A^*_{\ell \gamma} A\ns_{\ell\gamma} [\epsilon_\gamma(\zerov) - \mu] \\
{}+ \sum_{\{\ell\},\{\gamma\}}\bar{u}(\{\zerov\},\{\ell\},\{\gamma\}) A^*_{\ell_1\gamma_1}A^*_{\ell_2\gamma_2} A\ns_{\ell_3\gamma_3}A\ns_{\ell_4\gamma_4}\punc{,}
\end{multline}
where the interaction strength $\bar{u}$ is evaluated with all momenta equal to zero. [This expression has been divided by a factor of $(2\pi)^2\delta^2(\zerov)$, corresponding physically to the system volume.] The mean-field condensate configuration can be found by minimizing $h_0$ with respect to $A_{\ell\gamma}$. The resulting equation can be viewed as a time-independent Gross-Pitaevskii equation for the condensate wavefunction in momentum space.

The corresponding real-space wavefunction can be found using \refeqand{bFT}{bfromakl} and is given by
\beq{RealSpaceWavefunction}
\langle b_j \rangle = \sum_{\ell n} \omega^{n x_j + \ell y_j - n\ell} \sum_\gamma \psi_{\gamma n}(\zerov) A_{\ell\gamma}\punc{.}
\eeq
This is in general a function of $[x]_q$ and $[y]_q$, and so gives a $q \times q$ site unit cell in real space. To this order, the particle density is simply given by $\langle n_j \rangle = |\langle b_j \rangle|^2$. Note that the presence of $A_{\ell\gamma}$ for nonzero $\ell$ implies that the condensate contains components for $\kv = \ell \Y \neq \zerov$ and hence that spatial symmetry is broken. Within this mean-field theory, this spatial order develops simultaneously with the breaking of phase-rotation symmetry, and is a simple consequence of the degeneracy in $\ell$. (In fact, the finite-temperature transition in two dimensions is of the Berezinskii-Kosterlitz-Thouless type, and so this particular result is not necessarily reliable.)

\begin{figure}
\putinscaledfigure{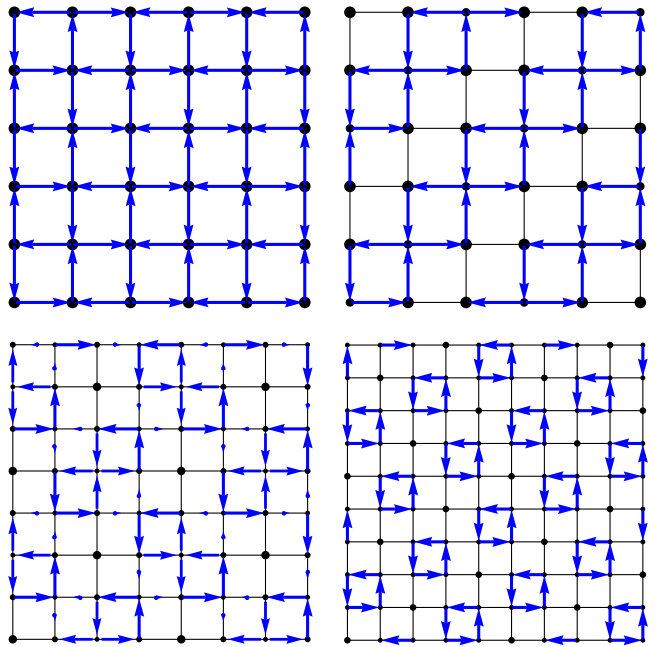}
\caption{\label{FigCurrents}Example mean-field condensate configurations for $\alpha = \frac{1}{2}$ (top left), $\frac{1}{3}$ (top right), $\frac{1}{4}$ (bottom left), and $\frac{1}{5}$ (bottom right). The blue arrows show the direction of the current $\langle \J_{ij} \rangle$ on each link $i \link j$ of the lattice, and their lengths indicate the magnitude (the length scale is not consistent between different values of $\alpha$). The black points show the positions of the lattice sites $i$ and have area proportional to the density $\langle n_i \rangle$. In each case, the condensate reduces the spatial symmetry of the square lattice, and is one member of a discrete set of degenerate configurations related by the action of the broken symmetries. The degeneracies and residual symmetries are listed in Table~\ref{TabConfigurations}. The quantitative details, but not the symmetries, depend on the interaction strength $U$; the plots show the case $U \ll t$.}
\end{figure}

The configuration of currents within the superfluid phase can be calculated using the gauge-invariant current operator for the link $i \link j$,
\beq{CurrentOperator}
\J_{ij} = \ii t \, \ee^{\ii \Phi_{ij}} b^\dagger_j b\nd_i + \text{h.c.}  \punc{.}
\eeq
Figure~\ref{FigCurrents} shows the currents $\langle\J_{ij}\rangle$ in the mean-field condensate configurations for $\alpha = \frac{1}{q}$ with $2 \le q \le 5$; they have the same $q\times q$ unit cell as the condensate wavefunction. For the larger values of $q$, particularly $\alpha = \frac{1}{5}$, these resemble Abrikosov lattices:\cite{Tinkham} the plaquettes with low density and high current circulation can be viewed as containing vortices. The symmetry properties of these configurations are listed in Table~\ref{TabConfigurations}.
\newcommand{\vs}{\;\;}
\begin{table}
\begin{tabular}{cccc}
$\alpha$ & $A_{\ell 1}$ & degeneracy & symmetries\\ \hline
$\frac{1}{2}$ & $(1\vs\ii)$ & $2$ & $\Ty\Tx$, $\Tx\R$, $\Ixy$\\
$\frac{1}{3}$ & $(1\vs\omega\vs\omega)$ & $6$ & $\Ty\Tx$, $\R^2$, $\Ixy$\\
$\frac{1}{4}$ & $\left(1\vs\sqrt{\omega/2}\vs{-\ii}\vs\sqrt{\omega/2}\right)$ & $16$ & $\R$, $\Ty^2\Tx^2\Iy$\\
$\frac{1}{5}$ & $\left(1\vs\omega\vs\omega^*\vs\omega^*\vs\omega\right)$ & $10$ & $\Ty\Tx^2$, $\R$
\end{tabular}
\caption{Properties of the mean-field condensate configurations shown in Figure~\ref{FigCurrents}, including their degeneracies and unbroken symmetries. The vectors in the column labeled $A_{\ell 1}$ are configurations minimizing $h_0$ in the limit of weak interactions, $U / t \rightarrow 0$, where the condensate is restricted to the lowest band, $\gamma = 1$. The residual symmetry group is given by products of powers of the operators listed, along with $\Tx^q$ and $\Ty^q$, which are always preserved by the ansatz of \refeq{BogvAnsatz}. The symmetries $\Tx$, $\Ty$, and $R$ are illustrated in Figure~\ref{FigLandauGaugeSymmetries}, and the combination $\Ixy = \Ix\R$ gives a reflection in the line $y = x$.}
\label{TabConfigurations}
\end{table}

An analysis of the patterns that are allowed for general interactions and various values of $q$ has been given by Balents et al.,\cite{Balents}\ who considered the same problem in a different context. In the present case, it is valid to assume purely on-site interactions, allowing the ordered states to be determined unambiguously.

Minimization of $h_0$ with respect to $A_{\ell\gamma}$ is equivalent to minimizing with respect to the real-space condensate wavefunction or vortex configuration. The latter perspective is more appropriate in the continuum, whereas here the lattice potential provides a strong pinning potential that simplifies the momentum-space approach.

Our ansatz for the condensate configuration, \refeq{BogvAnsatz}, also involves bands $\gamma$ other than the lowest. Occupation of higher bands costs kinetic energy, increasing the first term of \refeq{MFT}, and so is disfavored when interactions are very weak. For stronger interactions, the energy is reduced by smoothing out density fluctuations, which requires incorporating higher bands into the condensate. This competition between kinetic and potential energy also allows for first-order transitions between different local minima of $h_0$ as the interaction strength or mean density varies. We have not found any examples for $q \le 5$, however, and their observation in experiments would anyway likely require considerable enhancements in stability and cooling.

The mean-field energy $h_0$ is symmetric under the same transformations of $A_{\ell\gamma}$ as the full Hamiltonian is under transformations of $a_{\zerov \ell\gamma}$, as discussed in more detail in Appendix~\ref{AppSymmetries}. As noted above, certain symmetries are spontaneously broken by the condensate configuration, and the corresponding operators transform a given $A_{\ell\gamma}$ into a symmetry-equivalent degenerate configuration. The degeneracies of the patterns shown in Figure~\ref{FigCurrents} are listed in Table~\ref{TabConfigurations}. The number of degenerate configurations is in every case a multiple of $q$, as we prove in Appendix~\ref{AppSymmetries}. The degeneracy in the ordering patterns allows for the possibility of real-space domain formation, which would not affect time-of-flight images and would likely require more sophisticated {\it in situ} probes to confirm.\cite{Bakr,Sherson}

It should be noted that the ansatz of \refeq{BogvAnsatz} implicitly excludes ordered states with larger unit cells than $q \times q$ sites. (Previous work using a real-space approach\cite{Duric} has suggested that this may happen for $\alpha = \frac{1}{4}$.) It is straightforward to include such states (with larger but finite unit cells) at the mean-field level, by allowing nonzero condensate amplitude at a discrete set of momenta $[\kv]_{\BN} \neq \zerov$. This complicates somewhat the analysis that follows, and we will not treat this possibility further.

The condensate configuration $A_{\ell\gamma}$ also determines the occupation numbers in momentum space, and so can be used to predict the result of a time-of-flight expansion measurement, as discussed in detail in Section~\ref{SecTimeOfFlight}. Briefly, the terms in the Hamiltonian \refeq{Hamtk} mixing momenta differing by $\X$ imply Bragg peaks at points corresponding to momenta $n \X$, while the nonzero condensate amplitude $A_{\ell\gamma}$ for $\ell \neq 0$ gives further peaks at $\ell \Y$. (It should be recalled that our axes are reversed from those of Lin et al.\cite{Lin,LinNature})

\section{Bogoliubov theory}
\label{SecBogv}

The mean-field theory of Section~\ref{SecMFTheory} results from using the Bogoliubov ansatz of \refeq{BogvAnsatz} and keeping only the lowest-order term in an expansion in terms of the fluctuation operators $\at_{\kv\ell\gamma}$. To improve upon this theory and determine the spectrum for single-particle excitations above the condensate, we consider in this section the following order in the expansion. The terms containing a single operator vanish when the mean-field energy density $h_0$ is minimized, and so we next treat the quadratic terms.

The quadratic part of the Hamiltonian $\Ham^{(2)}$ can be conveniently expressed in matrix form, by combining creation and annihilation operators into a column vector,
\beq{Definealpha}
\alphav_{\kv\ell\gamma} =
\begin{pmatrix}
\at\nd_{\kv \ell \gamma}\\
\at^\dagger_{(-\kv) \ell \gamma}
\end{pmatrix}\punc{.}
\eeq
One can then write $\Ham^{(2)}$ as
\begin{multline}
\label{Quadratic}
\Ham^{(2)} = \frac{1}{2}\int_{\kv \in \BN}\!\dkv \sum_{\ell\ell',\gamma\gamma'}
\alphav^\dagger_{\kv\ell\gamma}
\Mm\nd_{\ell\gamma,\ell'\gamma'}(\kv)
\alphav\nd_{\kv\ell'\gamma'}\\
+ \Ham^{(2)}\sub{c}
\punc{,}
\end{multline}
where $\Ham^{(2)}\sub{c}$ is a term that contains no operators and comes from a commutator.

This expression can straightforwardly be generalized to allow for other choices of single-particle basis, by replacing $\ell$ and $\gamma$ by a single generic index $\lambda$. The matrix $\Mm(\kv)$ can then be written as
\beq{QuadraticMatrix}
\Mm_{\lambda\lambda'}(\kv) = \Imatrix_2[\epsilon_{\lambda\lambda'}(\kv) - \mu\delta_{\lambda\lambda'}] + \Bm_{\lambda\lambda'}(\kv)\punc{,}
\eeq
where $\epsilon_{\lambda\lambda'}$, the generalization of $\epsilon_\gamma\delta_{\gamma\gamma'}\delta_{\ell\ell'}$, is not diagonal in the general case, and
\begin{widetext}
\beq{BMatrix}
\Bm_{\lambda\lambda'}(\kv) = \sum_{\lambda_1\lambda_2}
\begin{pmatrix}
4\bar{u}(\kv,\zerov,\kv,\zerov;\lambda,\lambda_1,\lambda',\lambda_2)A_{\lambda_1}^* A_{\lambda_2}\ns&
2\bar{u}(\kv,-\kv,\zerov,\zerov;\lambda,\lambda',\lambda_1,\lambda_2)A_{\lambda_1} A_{\lambda_2}\\
2\bar{u}^*(\kv,-\kv,\zerov,\zerov;\lambda',\lambda,\lambda_1,\lambda_2)A_{\lambda_1}^* A_{\lambda_2}^*&
4\bar{u}(-\kv,\zerov,-\kv,\zerov;\lambda',\lambda_1,\lambda,\lambda_2)A_{\lambda_1}^* A_{\lambda_2}\ns
\end{pmatrix}\punc{.}
\eeq
\end{widetext}

Similarly to $h_0$ in \refeq{MFT}, $\Mm(\kv)$ has contributions from both the kinetic and potential energy. The latter can be viewed as self-energy terms for the quasiparticles due to scattering with bosons in the condensate. They include `anomalous' processes in which a pair of condensed particles scatter from each other into an excited state and the reverse process where they return to the condensate. These result in the off-diagonal elements in \refeq{BMatrix}, giving terms in $\Ham^{(2)}$ that do not conserve the number of $\at_{\kv \ell \gamma}$ quanta.\cite{Bogoliubov,AGD} The standard Bogoliubov theory for zero magnetic field is recovered by taking $q = 1$, in which case the $\ell$ and $\gamma$ indices are redundant.

It is useful to consider $\Mm_{\ell\gamma,\ell'\gamma'}(\kv)$ as a $2q^2\times 2q^2$ matrix (for each $\kv$). Using the properties of $u$ given after \refeq{HamUa}, one can show that this matrix is hermitian. The zero-momentum limit $\Mm(\zerov)$ is the Hessian of the mean-field term $h_0$ (with respect to variations in $A_{\ell \gamma}$ and its conjugate) and so is a nonnegative-definite matrix. The single vanishing eigenvalue corresponds to the broken $\mathrm{U}(1)$ symmetry of $h_0$. For nonzero $\kv$, all eigenvalues of $\Mm(\kv)$ are strictly positive.

\subsection{Bogoliubov quasiparticles}
\label{SecBogvQuasiparticles}

To find the spectrum of quasiparticles, one must define a new set of annihilation and creation operators in terms of which $\Ham^{(2)}$ is diagonal. Momenta (referred to $\BN$) are not mixed in \refeq{Quadratic}, so the new operators are labeled by $\kv$, but since the condensate breaks symmetry under $\Ty$, $\ell$ is no longer a good quantum number. We therefore define annihilation operators for these modes as $d_{\kv\zeta}$, where $\zeta \in \{1,2,\ldots,q^2\}$. (In certain cases, there are unbroken translation symmetries, as shown in Table~\ref{TabConfigurations}. The states can then be labeled by the eigenvalues of the corresponding operators, as discussed in Appendix~\ref{AppSymmetries}.)

In order to preserve the bosonic commutation relations, $[d\nd_{\kv\zeta}, d^\dagger_{\kv'\zeta'}] = (2\pi)^2\delta^2(\kv-\kv')\delta_{\zeta\zeta'}$, the transformation between $\alphav_{\kv\ell\gamma}$ and $d_{\kv\zeta}$ must be symplectic;\cite{Blaizot} details are given in Appendix~\ref{AppBogoliubov}. In terms of the new operators, the quadratic part of the Hamiltonian is given by
\begin{multline}
\label{H2Bogv}
\Ham^{(2)} = \int_{\kv \in \BN}\!\dkv \sum_{\zeta} \xi_{\kv\zeta} \bigg[  d^\dagger_{\kv\zeta} d\nd_{\kv\zeta}\\
{} - (2\pi)^2\delta^2(\zerov) \sum_{\ell\gamma}{Y^\zeta_{\kv\ell\gamma}}^{\!\!\!*} Y^\zeta_{\kv\ell\gamma} \bigg]\punc{,}
\end{multline}
where
\beq{dfroma}
d_{\kv\zeta} = \sum_{\ell\gamma}\left[{X^\zeta_{\kv\ell\gamma}}^{\!\!\!*} \at\nd_{\kv\ell\gamma} - {Y^\zeta_{\kv\ell\gamma}}^{\!\!\!*} \at^\dagger_{(-\kv)\ell\gamma} \right]\punc{.}
\eeq
To this order, the system is therefore described by noninteracting Bogoliubov quasiparticles with annihilation operators $d_{\kv\zeta}$ and energies $\xi_{\kv\zeta} > 0$. Because of the off-diagonal elements in $\Mm(\kv)$, the quasiparticles are superpositions of particles and holes, with $X^\zeta_{\kv\ell\gamma}$ and $Y^\zeta_{\kv\ell\gamma}$ respectively giving these components.

\begin{figure}
\putinscaledfigure{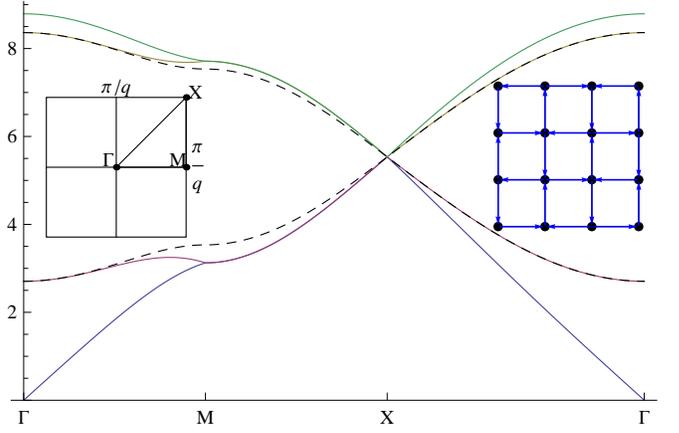}
\caption{\label{FigDispersions12}Quasiparticle dispersion (solid lines) and noninteracting single-particle dispersion (dashed), both in units of hopping $t$, for $\alpha = \frac{p}{q} = \frac{1}{2}$. An analytic expression for the spectrum for this case is given in \refeq{Eqxi2} of Appendix~\ref{AppAnalytics}. The dispersions are plotted along a path in the reduced Brillouin zone $\BN$ shown in the left inset. In the interacting case, $U = 4t$, the mean density is $\rho = 1$, and the real-space configuration is as shown in the right inset (see also Figure~\ref{FigCurrents}). In both cases there are $q^2 = 4$ modes, including, in the interacting case, one Goldstone mode with linear dispersion. For $U = 0$, the modes are $q$-fold degenerate and have been shifted vertically by an arbitrary choice of chemical potential. At $\kv = \frac{\pi}{2}\xhv + \frac{\pi}{2}\yhv$, the corner X of $\BN$, the modes meet at a point, with a linear dispersion. Such a `Dirac cone' occurs whenever $q$ is even. Other notable features of the interacting spectrum include a twofold degeneracy along the line from M to X, and the unshifted modes (relative to the noninteracting dispersion) from X to $\Gamma$. The former can be understood as a Kramers degeneracy due to the antiunitary symmetry under $\Tx\Iy$, as discussed in Appendix~\ref{AppSymmetries}.}
\end{figure}

\begin{figure}
\putinscaledfigure{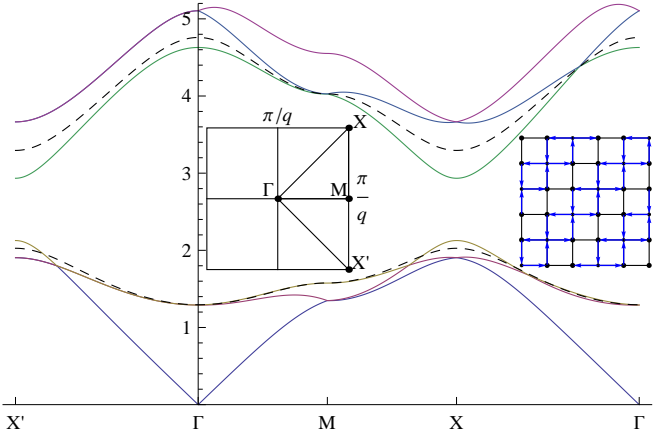}
\caption{\label{FigDispersions13}Quasiparticle dispersion (solid lines) and noninteracting single-particle dispersion (dashed), for $\alpha = \frac{p}{q} = \frac{1}{3}$. In the interacting case, $U = 2t$, the mean density is $\rho = 1$. In both cases, there are $q^2 = 9$ modes, of which only the lowest $6$ are shown. The dispersions are plotted along a path in the reduced Brillouin zone $\BN$ shown in the left inset. Because the condensate configuration breaks the symmetry under $\R$, rotation by $\frac{\pi}{2}$, the quasiparticle dispersions along the lines from $\Gamma$ to X and from $\Gamma$ to X$'$ are different. The interacting spectrum includes several (unavoided) level crossings, for example in the lower band near X$'$, a result of the unbroken translation symmetry $\Ty\Tx$, as discussed in Appendix~\ref{AppSymmetries}.}
\end{figure}

For vanishing interactions, the second term in \refeq{QuadraticMatrix} is absent and the quasiparticle spectrum $\xi_{\kv\zeta}$ is identical to the single-particle spectrum $\epsilon_\gamma(\kv)$ described in Section~\ref{SecSpectrum}. With nonzero interactions, the $q$-fold degeneracy within each band is split and, for generic $\kv$, the spectrum consists of $q^2$ distinct modes. The quasiparticle dispersion for $\alpha = \frac{1}{2}$ and $\frac{1}{3}$ are shown in Figures~\ref{FigDispersions12} and \ref{FigDispersions13} respectively, along with the noninteracting single-particle spectrum. (For clarity, only $6$ out of the $q^2 = 9$ modes are shown in the latter case.) For general $q$, the diagonalization of $\Mm(\kv)$ must be performed numerically, but the simplest case is analytically tractable, and is treated in detail in Appendix~\ref{AppAnalytics}.

As the figures show, the lowest energy approaches zero in the limit $\kv \rightarrow \zerov$, giving the Goldstone mode that results from broken $\mathrm{U}(1)$ symmetry. For small $|\kv|$ this `phonon' has a linear dispersion, and can be described in terms of long-wavelength fluctuations of the condensate phase. A low-energy theory of the mode can be found by allowing gradual deviations from the mean-field value of the phase and its conjugate density, as described in Section~\ref{SecAmplitudePhase}.

The phase velocity $c$ of the Goldstone mode, which is expressed in terms of the spectrum at $\kv = \zerov$ in Section~\ref{SecNearZeroMomentum}, is in many cases independent of direction, including both $\alpha = \frac{1}{2}$ and $\frac{1}{3}$.  This isotropy is a straightforward result of the symmetry in both cases under $\Ixy = \R\Iy$, as noted in Table~\ref{TabConfigurations}. This property, and other features of the spectra shown in Figures~\ref{FigDispersions12} and \ref{FigDispersions13}, are discussed in Appendix~\ref{AppSymmetries}.

The second term in \refeq{H2Bogv} includes $\Ham^{(2)}\sub{c}$ from \refeq{Quadratic} and represents the change in the zero-point energy associated with the superfluid state. This term, coming from quantum fluctuations, is accompanied at nonzero temperature $T$ by a contribution from thermally excited Bogoliubov quasiparticles, leading to a free energy per site of
\beq{FreeEnergyCorrection}
\Delta f = \int_{\kv\in\BN}\!\dkv \sum_\zeta \bigg[ T\log(1-\ee^{-\xi_{\kv\zeta}/T})
 - \xi_{\kv\zeta} |Y^\zeta_{\kv\ell\gamma}|^2 \bigg]\punc{,}
\eeq
where we use units such that $k\sub{B} = 1$. (Within the mean-field theory of Section~\ref{SecMFTheory}, all particles are in the condensate, so the entropy vanishes and the free energy is given by $h_0$.) These contributions in principle allow a configuration with a higher mean-field energy $h_0$ to be selected because of its enhanced fluctuations and hence lower free energy $h_0 + \Delta f$. It should be noted, however, that the degeneracy of the symmetry-equivalent condensate configurations discussed in Section~\ref{SecMFTheory} cannot be lifted by $\Delta f$.

The calculated spectra lead to important experimental predictions, as discussed below in Section~\ref{SecExperiment}. Occupation of the quasiparticle modes, due to both thermal and quantum fluctuations, gives the structure of time-of-flight images away from the Bragg peaks mentioned previously, and spectroscopic methods should be able to measure the mode dispersions $\xi_{\kv\zeta}$ directly (see Section~\ref{SecSpectroscopy}).

\subsection{Condensate depletion}
\label{SecDepletion}

The ansatz of \refeq{BogvAnsatz} and the expansion in powers of operators is in principle exact, with the higher-order terms leading to interactions between the Bogoliubov quasiparticles. Here, the series is truncated at quadratic order, an approximation that is valid provided that the quasiparticles remain at sufficiently low density for their interactions to be neglected.

This criterion can be quantified by calculating the depletion of the condensate, equal to the quasiparticle contribution to the total particle number. This is found by expressing the number operator $n_j$ is terms of the quasiparticle operators $d_{\kv\zeta}$, summing over sites $j$, and taking the ensemble average. The mean particle density is then given by
\begin{multline}
\label{DepletionDensity}
\rho = \sum_{\ell\gamma} |A_{\ell\gamma}|^2 + \int_{\kv\in\BN}\!\dkv \sum_{\zeta\ell\gamma} \bigg\{ |X^\zeta_{\kv\ell\gamma}|^2 n\sub{B}(\xi_{\kv\zeta}) \\
{} + |Y^\zeta_{\kv\ell\gamma}|^2 [1 + n\sub{B}(\xi_{\kv\zeta})] \bigg\} \punc{,}
\end{multline}
where $n\sub{B}(\xi) = (\ee^{\xi/T} - 1)^{-1}$ is the Bose-Einstein distribution function. The first term in \refeq{DepletionDensity} is the condensate density, and is simply the spatial average of the mean-field density calculated in Section~\ref{SecMFTheory}, while the second term gives the average density of particles outside the condensate. The relative magnitude of these two terms gives a measure of the significance of fluctuations, and we use the ratio of the second to the first as our definition of the depletion.

\begin{figure}
\putinscaledfigure{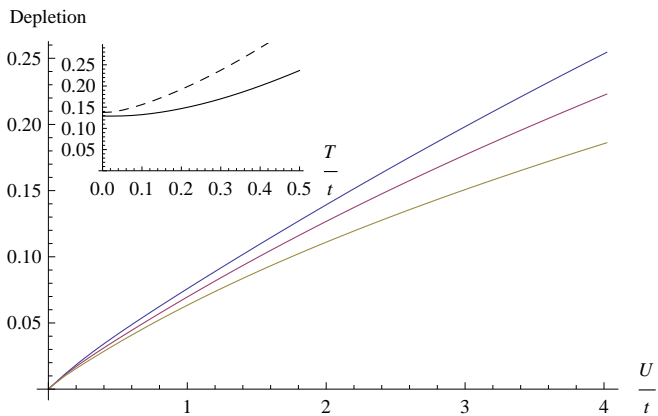}
\caption{\label{FigDepletion}(Color online) Condensate depletion for $\alpha = \frac{1}{3}$ as a function of interactions $U/t$ (main figure) and temperature $T/t$ (inset), where $t$ is the hopping strength. In the main figure, $T = 0$ and the densities are $\rho = 1$ (top curve), $2$ (middle), and $4$ (bottom), while in the inset, $\rho = 1$ and $U/t = 2$. The depletion is smallest, and hence the approximation best, deep in the superfluid phase, with weak interactions, high density, and low temperature. For $T>0$, small-momentum cutoffs of $k_0 = 0.1$ (solid line) and $k_0 = 0.02$ (dashed line), in lattice units, have been used to remove the logarithmic divergence of the depletion integral.}
\end{figure}
Figure~\ref{FigDepletion} shows the depletion for $\alpha = \frac{1}{3}$, as a function of density, interaction strength, and (in the inset) temperature. It is small deep within the superfluid phase and increases to roughly $25\%$ for the largest values of $U$ and $T$ shown. Neglecting cubic and quartic terms within the Bogoliubov theory relies on the assumption of small depletion, and so the conclusions presented here are only qualitatively applicable for larger values of $U$ and $T$. (The order of magnitude is consistent with the spin-wave analysis of \DJ uri\'c and Lee.\cite{Duric})

For zero temperature, $n\sub{B}(\xi > 0) = 0$ and the only fluctuation contribution is from the second term within the braces [zero-point fluctuations; compare \refeq{FreeEnergyCorrection}]. In this case, the integrand diverges as $v_0^2|\kv|^{-1}$ for small $|\kv|$, where $v_0$ is a coefficient in the expansion of $Y_{\kv\ell\gamma}^\zeta$ for small $\kv$ (see Appendix~\ref{SecNearZeroMomentum}). The integral is therefore finite in this case.

For $T>0$, the Bose-Einstein distribution function becomes $T(c|\kv|)^{-1}$ for small $|\kv|$. This results in a logarithmically divergent integral, an instance of the Mermin-Wagner-Hohenberg theorem,\cite{Mermin,Hohenberg} which states that, in two dimensions for nonzero temperature, the continuous phase symmetry cannot be broken. In an infinite two-dimensional system, there is no true condensate and so the `depletion' is complete.

In the presence of an external trapping potential, however, nonzero temperature condensation is possible even in two dimensions.\cite{Bagnato} This can be captured in a crude way by applying a small-momentum cutoff $k_0$ on the integral over $\kv$, with $k_0 \simeq R\sub{eff}^{-1}$, where $R\sub{eff}$ is the effective radius of the system in the trap (in units of the lattice spacing).

If the two-dimensional plane is embedded within a deep lattice in the $z$ direction, then hopping in this transverse direction can also stabilize the condensate. An appropriate momentum cutoff is then given by $k_0 \approx \sqrt{2m^* t_{\perp}}$, where $t_{\perp}$ is the transverse hopping matrix element, and hence the energy scale over which the system appears three-dimensional, and $m^*$ is the effective mass at the minimum of the lowest band in the single-particle dispersion.

In either case, the depletion integral is finite, with a logarithmic dependence on $k_0$ of
\beq{DepletionLogTerm}
\rho\sub{log} = \frac{v_0^2 T}{2\pi c} \log k_0\punc{.}
\eeq
In the inset of Figure~\ref{FigDepletion}, the depletion is shown at nonzero temperature, using two different values of $k_0$. The difference between the two curves is well approximated for most values by \refeq{DepletionLogTerm}. (The exact difference involves other terms that are not singular at $k_0 = 0$.) To determine the cutoff used in the plot, we have assumed that the finite system size will be the most important effect, and taken $\R\sub{eff}\simeq 10$--$50$ lattice sites.\cite{JimenezGarcia}

The depletion calculation also provides a rough estimate for the boundary of the superfluid phase, at the point where the depletion reaches $100\%$, although the approximation of independent quasiparticles is probably not valid at this point. For $\alpha = \frac{1}{3}$, $\rho = 1$ and $T = 0$, this gives an estimate of $(t/U)\sub{c} = 0.08$, in reasonable agreement with the value of $(t/U)\sub{c} = 0.063$ (at the tip of the $\rho = 1$ Mott lobe) found using the Gutzwiller ansatz.\cite{Oktel,Umucalilar,Goldbaum} It should be noted that the latter approach, which neglects fluctuations within the Mott insulator, generally underestimates $(t/U)\sub{c}$.\cite{CapogrossoSansone}

\section{Experimental predictions}
\label{SecExperiment}

The Bogoliubov theory that we have presented for the superfluid phase provides several concrete predictions for experiments, most notably for time-of-flight images and Bragg spectroscopy.

\subsection{Time-of-flight images}
\label{SecTimeOfFlight}

As noted above in Section~\ref{SecMFTheory}, the enlarged unit cell of the condensate has important consequences for time-of-flight images. The corresponding reduction of the Brillouin zone leads to additional Bragg peaks that give a clear indication of the formation of spatial order in the condensate. The intensity away from these peaks is determined by bosons excited to states with $[\kv]_{\BN}\neq \zerov$ by thermal and quantum fluctuations.

In a time-of-flight measurement, the trapping potential confining the atoms within the lattice is suddenly switched off, causing a rapid expansion. After a fixed period of the time, the density profile of the cloud is determined, for example by illuminating the atoms and measuring the transmitted intensity. If the interactions between the atoms during the expansion are sufficiently weak, then it can be treated as ballistic, and we will assume that this is the case throughout. In the absence of a magnetic field, the density profile after a fixed time of flight measures the original momentum distribution in the trap.\cite{Toth} The same is true with a field, apart from some modifications that we discuss in the following.

The time-of-flight images depend on certain details of the experiment and, in particular, the means used to produce the effective magnetic field. In the case of a rotating system, the momentum in the stationary (laboratory) frame is equal, up to a possible global rotation, to the symmetric-gauge canonical momentum in the rotating frame.

With a Raman-induced gauge field, the results depend on whether the Raman beams remain after release; we assume that they are suddenly switched off simultaneously with the trap, such as in the experiments of Lin et al.\cite{LinNature} The trajectory of an atom is determined by the momentum immediately after the gauge field is switched off, which is equal, using the sudden approximation, to the Landau-gauge canonical momentum before switch-off.\cite{Footnote1}

Within the approximation of a ballistic expansion, the time-of-flight image shows the continuum momentum occupation $N(\kv)$, defined by
\beq{ContinuumMomentumOccupation}
N(\kv) = \langle \Psit^\dagger(\kv)\Psit(\kv) \rangle\punc{,}
\eeq
where $\Psit(\kv)$ is the (continuum) momentum-space annihilation operator.\cite{Toth} The real-space operator $\Psi(\rv)$ can, after projection to the lowest Bloch band, be expressed in terms of the lattice operator $b_j$ using the Wannier function $W_j(\rv)$,
\beq{ContinuumToLattice}
\Psi(\rv) = \sum_j W_j(\rv) b_j\punc{.}
\eeq
In the presence of a magnetic field, this expression cannot generally be written as a convolution.

Combing \refeqand{ContinuumMomentumOccupation}{ContinuumToLattice} and using \refeq{bFT} to express $b_j$ in terms of $b_{\kv}$ gives
\beq{CMOfromF}
N(\kv) = \int_{\kv_1,\kv_2 \in \BL} F^*(\kv,\kv_1) \langle b_{\kv_1}^\dagger b_{\kv_2}\nd\rangle F(\kv,\kv_2)
\eeq
(where, for brevity, the standard integration measure for both integrals has been omitted). The kernel of this double integral transform, analogous to a matrix similarity transformation, is given by
\beq{KernelF}
F(\kv,\kv') = \sum_j \ee^{\ii \kv'\cdot \xv_j} \int \dd^3\rv\, \ee^{-\ii \kv\cdot \rv} W_j(\rv)\punc{,}
\eeq
where $\rv$ is integrated over all space. (The Wannier function $W_j$ restricts the integral to the neighborhood of the two-dimensional plane.)

The kernel $F$ depends on experimental details, including the optical lattice parameters and the effective gauge potential, while theoretical analysis based on the Bose-Hubbard model leads to predictions for the correlation function $\langle b_{\kv_1}^\dagger b_{\kv_2}\nd\rangle$. We will first outline the form of $F$ appropriate to experiments using rotation and Raman-induced gauge fields, before giving our results for the correlation function based on the Bogoliubov theory.

In the absence of a magnetic field, the Wannier function at site $j$ is a function only of $\rv - \xv_j$, and so can be written in the form $W\zero_j(\rv) = w\zero(\rv - \xv_j)$. Shifting the integration variable $\rv$ in \refeq{KernelF} allows the sum to be evaluated, giving
\beq{KernelF0}
F\zero(\kv,\kv') = (2\pi)^2\delta^2([\kv - \kv']_{\BL}) \tilde{w}\zero(\kv)\punc{,}
\eeq
where $\tilde{w}\zero$ is the Fourier transform of $w\zero$. This leads to the simple result
\beq{Nzero}
N\zero(\kv) = |\tilde{w}\zero(\kv)|^2 \left\langle b^\dagger_{[\kv]_{\BL}} b\nd_{[\kv]_{\BL}} \right\rangle\punc{,}
\eeq
so the time-of-flight image gives the lattice-momentum distribution, with an overall envelope given by the Wannier function.\cite{Toth}

With nonzero gauge potential $\scAv$, one can instead express the Wannier function as\cite{Luttinger,Wannier}
\beq{Wannier}
W_j(\rv) = w(\rv - \xv_j)\exp\left[\ii\int_{\xv_j}^{\rv} \dd\rv' \cdot \scAv(\rv')\right]\punc{,}
\eeq
where the integral is taken along a straight-line path, as in \refeq{HoppingPhases}. In this case, the kernel $F$ is no longer diagonal in $\kv$ and $\kv'$, and will depend on the appropriate choice of gauge.

In the Landau gauge (with $\scAv$ parallel to $\yhv$), one can write $\scAv\sub{L}(\rv) = \scBv \times r_x\xhv$, and the kernel is given by
\begin{multline}
\label{FLandau}
F\sub{L}(\kv,\kv') = \int \dd^3 \rv \, \ee^{-\ii \kv \cdot \rv} w(\rv) \ee^{\frac{1}{2}\ii |\scBv| r_x r_y}\\
\times (2\pi)^2\delta^2([\kv - \kv' + \scBv \times r_y\yhv ]_{\BL})\punc{.}
\end{multline}
In the symmetric gauge, $\scAv\sub{S}(\rv) = -\frac{1}{2}\scBv \times \rv$, leading to
\begin{multline}
\label{FSymm}
F\sub{S}(\kv,\kv') = \int \dd^3 \rv \, \ee^{-\ii \kv \cdot \rv} w(\rv) \\
\times (2\pi)^2\delta^2([\kv - \kv' + \frac{1}{4}\scBv \times \rv]_{\BL})\punc{.}
\end{multline}
It should be noted that, in both cases, $w(\rv)$ differs from the Wannier function $w\zero(\rv)$ in the absence of a magnetic field.

For our purposes, it is sufficient to note that the kernels both give contributions to $N(\kv)$ from a range of lattice momenta $[\kv]_{\BL} + \delta\kv$. The scale is determined by $|\delta\kv| \lesssim |\scBv| d = 2\pi \alpha d$, where $d$ is the characteristic size of the Wannier function $w$. This point-spreading effect can be understood as resulting from the position-dependent impulse $\scAv(\rv)$ imparted to the atoms when the gauge potential is switched off.

We now discuss the form of the correlation function $\langle b_{\kv_1}^\dagger b\nd_{\kv_2} \rangle$, which also depends on the gauge. As in previous sections, we will focus on the Landau gauge, appropriate for experiments with Raman-induced gauge potentials.

The symmetry under $\Tx^q$ and $\Ty^q$ implies that $\langle b_{\kv_1}^\dagger b\nd_{\kv_2} \rangle$ vanishes unless $[\kv_1 - \kv_2]_{\BN} = \zerov$. We therefore define
\beq{NkDefine}
(2\pi)^2\delta^2(\kv - \kv') N_{n\ell,n'\ell'}(\kv) = \langle b^\dagger_{\kv+n \X + \ell \Y} b\nd_{\kv'+n' \X + \ell' \Y} \rangle\punc{,}
\eeq
for $\kv,\kv' \in \BN$. (The formally infinite factor when $\kv = \kv'$ corresponds physically to system volume.)

The dominant contribution to $N_{n\ell,n'\ell'}(\kv)$ is a delta-function peak at $\kv = \zerov$, coming from the first term in \refeq{BogvAnsatz},
\beq{NkLandau}
N_{n\ell,n'\ell'}(\kv) = (2\pi)^2\delta(\kv)\sum_{\gamma\gamma'} A^*_{\ell\gamma} \psi^*_{\gamma n}(\kv)A\ns_{\ell'\gamma'}\psi\ns_{\gamma' n'}(\kv)\punc{.}
\eeq
The corresponding peaks in $\langle b_{\kv_1}^\dagger b\nd_{\kv_2} \rangle$, at momenta such that $[\kv_1]_{\BN} = [\kv_2]_{\BN} = \zerov$, are separated by multiples of $\X$ and $\Y$. If they are to be resolved in time-of-flight images, we require that their separation, $2\pi/q$, be greater than the point-spread $2\pi\alpha d$ of the kernel $F$. The condition is then simply that the Wannier function $w(\rv)$ be well localized compared to the lattice spacing.

If this condition is satisfied, then the time-of-flight intensity $N(\kv)$ consists of sharp peaks near each of the reciprocal lattice vectors $n \X + \ell \Y$ of the reduced Brillouin zone $\BN$. These extra Bragg peaks in fact result from two separate physical effects.

The first is the enlargement of the unit cell of the Hamiltonian to $q \times 1$ sites, as a result of the phases appearing the hopping term $\Ham_t$. States with momentum differing by $\X$ are therefore mixed at the single-particle level, leading to additional Bragg peaks at momenta $2 \pi \xhv n / q$ even in the absence of interactions.\cite{Gerbier} The observation of such peaks in an experiment is a clear sign that the flux per plaquette is at (or sufficiently close to) a rational value.

The second effect occurs only in the presence of interactions and is due to the spontaneous breaking of spatial symmetry in the superfluid. As discussed in Section~\ref{SecMFTheory}, the condensate contains contributions from the $q$ degenerate minima of the single-particle dispersion, and therefore enlarges the unit cell to $q \times q$ sites. Peaks at momenta $2 \pi \yhv \ell / q$ are clear indications of the formation of such an ordered state.

Importantly, the kernel for the Landau gauge, $F\sub{L}(\kv,\kv')$ in \refeq{FLandau}, does not change the $y$ component of the momentum, and so the point-spreading effect is entirely in the $x$ direction. The second class of Bragg peaks, resulting from interaction effects, are therefore not affected, increasing the likelihood that they can be observed in experiment. Note that this separation does not apply in the symmetric gauge, appropriate for the case of rotation, and furthermore that the spacing of the Bragg peaks is reduced, as a result of the $2q\times 2q$ unit cell of $\Ham_t$, making their observation considerably more challenging.

Finally, it should be noted that in many cases, including $\alpha = \frac{1}{2}$ and $\frac{1}{3}$, the condensate has equal amplitude (but not phase) for all values of $\ell$. Bragg peaks with the same value of $n$ therefore have the same intensity, apart from the envelope coming from the on-site Wannier wavefunction. This is in contrast to the case of strictly vanishing interactions, when any distribution of particles between the $q$ minima of the single-particle dispersion has equal probability.

\subsection{Spectroscopy}
\label{SecSpectroscopy}

Developments in spectroscopic measurements for ultracold atomic systems\cite{Stenger,StamperKurn,Stewart,Kollath,Sensarma,Strohmaier,Clement} have allowed experimental access to dynamic correlation functions within these systems. We consider two such techniques, Bragg spectroscopy\cite{Stenger,StamperKurn,Rey} and lattice-modulation spectroscopy,\cite{Sensarma,Strohmaier} and describe the information regarding the quasiparticle spectrum that can be determined from both.

Bragg spectroscopy\cite{Stenger,StamperKurn,Rey} involves applying a weak periodic perturbation of the form $\cos (\Kv \cdot \xv - \Omega t)$ to the system, using two laser beams at an angle and with frequencies differing by $\Omega$. One then measures, usually through time-of-flight imaging, the total momentum or energy imparted to the system. The response is given by the dynamical structure factor $S(\Kv,\Omega)$, the density correlation function in momentum and frequency space.

Lattice-modulation spectroscopy\cite{Sensarma,Strohmaier} involves oscillating the lattice depth at frequency $\Omega$; using time-of-flight imaging to determine the imparted energy then gives $S(\zerov,\Omega)$. It is also possible to measure $S(\Kv,\Omega)$ at certain high-symmetry points $\Kv$ in the lattice Brillouin zone $\BL$ by the application of lattices with enlarged periods. (Each point in $\BL$ corresponds to a point in $\BN$ in a way that depends on $q$.) Even if only the point $\Kv = \zerov$ is accessible, the presence of multiple Hofstadter bands should be clear, and the splitting of the Goldstone mode from the rest of the first band is also measurable.

In either case, the coupling to the perturbation can be expressed in terms of the momentum-space density operator,
\beq{Definerhok}
\rho(\Kv) = \int_{\kv\in\BL}\dkv b^\dagger_{\kv - \Kv} b\nd_{\kv} = \sum_j \ee^{-\ii \xv_j \cdot \Kv} n_j\punc{,}
\eeq
which, being a function only of $n_j$, is gauge-invariant. The dynamic structure factor is given in spectral representation by
\begin{multline}
\label{DSF}
S(\Kv,\Omega) = \frac{1}{\Z} \sum_{\Psi_1,\Psi_2} \ee^{-E_{\Psi_1}/T} \delta(E_{\Psi_2} - E_{\Psi_1} - \Omega)\\
\times {|\langle \Psi_2 | \rho(\Kv) | \Psi_1 \rangle|}^2\punc{,}
\end{multline}
where $\Z = \sum_\Psi \ee^{-E_\Psi/T}$ is the partition function, and $|\Psi_{1,2}\rangle$ are eigenstates of the Hamiltonian $\Ham$ with energy $E_{\Psi_{1,2}}$.

While $S(\Kv,\Omega)$ is given by a four-point correlation function, it can be factorized into two-point functions within the quadratic Bogoliubov theory. In the condensed phase, and assuming depletion is not too large, the dominant contribution to the integral in \refeq{Definerhok} in fact comes from the points where either $[\kv]_{\BN} = \zerov$ or $[\kv - \Kv]_{\BN} = \zerov$. (For $[\Kv]_{\BN} = \zerov$ these cases coincide, and there is an extra term in $\rho(\Kv)$ which, however, contributes only at $\Omega = 0$.) The structure factor is therefore given by a two-point correlation function multiplied by the condensate density.

Within this approximation, the density operator can be expanded in terms of the operators $d_{\kv\zeta}\nd$ and $d_{\kv\zeta}^\dagger$:
\begin{widetext}
\beq{rhofromd}
\rho([\Kv + N\X + L\Y]_{\BL}) = \sum_\zeta \left\{r^\zeta_{NL}(\Kv)d_{\Kv\zeta}\nd + \omega^{-2y_0 L}[r^\zeta_{NL}(\Kv)]^* d_{-\Kv\zeta}^\dagger\right\}\punc{,}
\eeq
where
\begin{multline}
r^\zeta_{NL}(\kv) = \sum_{n,l=0}^{q-1} \sum_{\gamma \gamma'} \Big[\omega^{-n\ell+(n-N)(\ell-L)}A^*_{[\ell-L]_q\gamma}\psi^*_{\gamma[n-N]_q}(\zerov)\psi\ns_{\gamma'n}(\kv)X^\zeta_{\kv\ell\gamma'}\\
{}+ \omega^{n\ell-(n+N)(\ell+L)}A\ns_{[\ell+L]_q\gamma}\psi\ns_{\gamma[n+N]_q}(\zerov)\psi^*_{\gamma'n}(-\kv)Y^\zeta_{\kv\ell\gamma'}\Big]\punc{.}
\end{multline}
A subdominant third term involving two $d_{\kv\zeta}$ operators has been dropped from \refeq{rhofromd}.

The dynamic structure factor defined in \refeq{DSF} is calculated using the eigenstates of $\Ham$, which, at the level of the quadratic approximation of Section~\ref{SecBogv}, are eigenstates of the occupation numbers of each Bogoliubov mode. The matrix elements of \refeq{rhofromd} between any pair of states can be expressed in terms of these occupation numbers, giving
\beq{EqFinalDSF}
S([\Kv + N\X + L\Y]_{\BL},\Omega) = \sum_\zeta |r_{NL}^\zeta(\Kv)|^2 \Big\{\delta(\Omega + \xi_{\Kv\zeta}) n\sub{B}(\xi_{\Kv\zeta}) \\
{}+ \delta(\Omega - \xi_{\Kv\zeta})[1+n\sub{B}(\xi_{\Kv\zeta})]\Big\}\punc{.}
\eeq
\end{widetext}
The structure factor at frequency $\Omega$ therefore has resonances at each quasiparticle mode $\zeta$, allowing the quasiparticle spectrum to be measured directly.

\section{Discussion}
\label{SecDiscussion}

We have studied the effect of a synthetic magnetic field on the superfluid phase of bosons in a lattice. Our theoretical approach is based on Bogoliubov theory, which determines the condensate configuration and allows interactions to be taken into account within an expansion in terms of fluctuations. We predict broken spatial symmetry in the condensed phase, leading to qualitative changes compared to the Hofstadter spectrum for noninteracting particles.

This analysis leads to several clear predictions that should be testable in experiment. The density modulations in the superfluid phase, illustrated in Figure~\ref{FigCurrents}, may be directly measurable using recently developed real-space imaging techniques.\cite{Bakr,Sherson} Our order-of-magnitude estimate for the extent of superfluidity given in Section~\ref{SecDepletion}, which is in agreement with independent theoretical approaches,\cite{Oktel,Umucalilar,Goldbaum} can be tested in experiments analogous to those performed in the absence of a magnetic field.\cite{Greiner} Predictions for spectroscopic measurements have been detailed in Section~\ref{SecSpectroscopy}.

Our approach also provides predictions for time-of-flight imaging, the most well-established technique in cold-atom experiments. As described in Section~\ref{SecTimeOfFlight}, we predict extra Bragg peaks due to the spatial symmetry breaking. In experiments using Raman-induced gauge fields, these result from two distinct physical effects. The gradient in the applied synthetic vector potential (due to a gradient in the physical magnetic field in the experiments of Lin et al.\cite{LinNature}) breaks translation symmetry explicitly, leading to an extra set of Bragg peaks in the direction of the gradient. By contrast, symmetry under translation in the perpendicular direction is broken spontaneously when the bosons condense, and this leads to further Bragg peaks, in the direction of propagation of the applied Raman lasers ($\xhv$ in our convention, but $\yhv$ in the experiments). The appearance of these latter peaks is therefore a clear signature of many-body effects.

Among the approximations made in the present work is the assumption that thermal equilibrium can be reached on the time scale of the experiments. Previous studies of closely related systems\cite{Goldbaum2} have shown that the process of vortex formation can exhibit hysteresis, and experiments with effective gauge potentials exhibit a considerable dependence of the vortex density on hold times.\cite{LinNature} In the model considered here, two-body scattering is sufficient to populate modes of nonzero $\ell$ and hence generate nontrivial spatial structures, but further work is required to provide quantitative estimates of the rate for these processes.

We have also neglected the influence of higher lattice bands and hopping between pairs of sites other than nearest neighbors. Neither is expected to have qualitative effects on our conclusions, as long as the magnetic symmetries described in Section~\ref{SecMSG} are preserved. (This is certainly the case with a synthetic magnetic field due to rotation or Raman lasers, but not necessarily so when hopping phases are induced by other methods.\cite{Dum,JakschZoller,Mueller,Sorensen,Gerbier}) As already noted in Section~\ref{SecHamiltonian}, weak interactions between bosons on different sites will also have only quantitative effects.

As discussed in Section~\ref{SecMFTheory}, the broken spatial symmetry implies the existence of multiple degenerate configurations in the superfluid phase. This allows for the possible formation of real-space domains, especially on shorter time scales, upon which the effect of the external trapping potential is likely to be important.

Besides the simplifications inherent in our starting model, our analysis has made the approximation of truncating the Bogoliubov expansion at quadratic order, neglecting interactions between quasiparticles. Consequences of these interactions include finite quasiparticle lifetimes and also the possibility of spectrum termination at the point where decay into the two-particle continuum is allowed by kinematics. These are likely to have implications for spectroscopy experiments; an understanding of these is left for future work.

\begin{acknowledgments}
We thank Ian Spielman, Trey Porto, Chris Foot, Ross Williams, and Sarah Al-Assam for helpful discussions. This work is supported by JQI-NSF-PFC, ARO-DARPA-OLE, and Atomtronics-ARO-MURI.
\end{acknowledgments}

\appendix

\section{Bogoliubov transformation}
\label{AppBogoliubov}

In this Appendix, we will show that to diagonalize $\Ham^{(2)}$, given in \refeq{Quadratic}, one must find the eigenvalues and -vectors of the matrix $\etam \Mm_{\kv}$ for each $\kv \in \BN$, where
\beq{Defineeta}
\etam_{\ell\gamma,\ell'\gamma'} = \delta_{\ell\ell'}\delta_{\gamma\gamma'}
\begin{pmatrix}
1&0\\
0&-1
\end{pmatrix}\punc{,}
\eeq
and the matrix product is taken treating both $\etam$ and $\Mm_{\kv}$ as $2q^2 \times 2q^2$ matrices. While $\etam \Mm_{\kv}$ is not hermitian, it can be shown\cite{Blaizot} that, since $\Mm_{\kv}$ is nonnegative-definite, the eigenvalues of $\etam \Mm_{\kv}$ are all real. Furthermore, for $\kv \neq \zerov$, when $\Mm_{\kv}$ is {\it positive}-definite, the eigenvalues are all nonzero and come in pairs with equal magnitude and opposite sign. In this case, the $q^2$ positive eigenvalues $\xi_{\kv\zeta}$ and the corresponding eigenvectors $\Vv_{\kv}^\zeta$,
\beq{DefineV}
V_{\kv\ell\gamma}^\zeta =
\begin{pmatrix}
X_{\kv\ell\gamma}^\zeta\\
Y_{\kv\ell\gamma}^\zeta
\end{pmatrix}\punc{,}
\eeq
defined by $\etam \Mm_{\kv} \Vv_{\kv}^\zeta = \xi_{\kv \zeta} \Vv_{\kv}^\zeta$, describe the Bogoliubov quasiparticles.

Section~\ref{SecNearZeroMomentum} treats separately the special case of $\kv = \zerov$, allowing us to develop a series expansion for the properties near this point.

\subsection{Inversion symmetry}
\label{SecInversionSymmetry}

The Bogoliubov transformation, which mixes annihilation operators at momentum $\kv$ with creation operators at $-\kv$, requires the existence of an inversion symmetry in the condensed phase. Because of broken translational symmetry in the presence of a condensate, it is not necessarily the case that $\P$, inversion about the origin $\xv = \zerov$, remains a good symmetry. Instead, define the operator $\P_{y_0} = \Ty^{2y_0} \P$ representing inversion about the real-space point $\xv = y_0 \yhv$ (a lattice site if $y_0$ is an integer or the center of a bond if $y_0$ is a half integer).

Using \refeqand{Tyakl}{Pak}, $\P_{y_0}$ can be shown to obey
\beq{Pyak}
\P_{y_0} a_{\kv \ell \gamma} = \ee^{-2\ii y_0 k_y}\ee^{-2\ii \phi_\gamma} \omega^{-2y_0 \ell} a_{(-\kv)[-\ell]_q \gamma} \P_{y_0}\punc{,}
\eeq
so the condensate is invariant under this transformation if $A_{\ell\gamma}$ obeys
\beq{CondensateInversionSymmetry}
A_{\ell\gamma} = A_{[-\ell]_q\gamma} \ee^{-2\ii\phi_\gamma} \omega^{-2y_0\ell}\punc{.}
\eeq
The following assumes that the condensate has an inversion point, and hence there exists some value of $y_0$ for which this relation holds.

Corresponding to this inversion symmetry, and analogous to the matrix $\etam$, define the $2q^2\times 2q^2$ matrix $\pim$,
\beq{Definetheta}
\pim_{\ell\gamma,\ell'\gamma'} = \delta_{[\ell+\ell']_q,0}\delta_{\gamma\gamma'}\ee^{-2\ii\phi_\gamma}
\begin{pmatrix}
\omega^{-2y_0\ell}&0\\
0&\omega^{2y_0\ell}
\end{pmatrix}\punc{.}
\eeq
It is also convenient\cite{Blaizot} to define $\gammam$,
\beq{Definegamma}
\gammam_{\ell\gamma,\ell'\gamma'} = \delta_{\ell\ell'}\delta_{\gamma\gamma'}
\begin{pmatrix}
0&1\\
1&0
\end{pmatrix}\punc{,}
\eeq
which has the effect of exchanging creation and annihilation operators: $\gammam \alphav_{\kv} = (\alphav^\dagger_{-\kv})^\tr$. (The matrices $\etam$, $\pim$, and $\gammam$ are all both hermitian and unitary, and obey $\etam\pim = \pim\etam$, $\pim\gammam=\gammam\pim^*$, and $\gammam\etam = -\etam\gammam$.)

Under the assumption that $A_{\ell\gamma}$ obeys \refeq{CondensateInversionSymmetry}, one can show that $\pim \Mm_{\kv} \pim = \Mm_{-\kv}$, and furthermore, $\gammam\pim\Mm\ns_{\kv}\pim\gammam = \Mm_{\kv}^*$. These symmetries imply that for every eigenvector $\Vv_{\kv}^\zeta$ of $\etam\Mm_{\kv}$ with a positive eigenvalue $\xi_{\kv\zeta}$, there is a corresponding eigenvector $\Wv_{\kv}^\zeta = \pim\gammam{\Vv_{\kv}^\zeta}^*$ with eigenvalue $-\xi_{\kv\zeta}$. The corresponding eigenvector of $\etam\Mm_{-\kv}$ is $\Wv_{-\kv}^\zeta = \gammam {\Vv_{\kv}^\zeta}^*$, so that $\xi_{-\kv\zeta} = \xi_{\kv\zeta}$.

\subsection{Nonzero momentum}
\label{SecBogvNonzero}

For nonzero $\kv$, the matrix $\Mm_{\kv}$ is positive-definite, and all eigenvalues of $\etam\Mm_{\kv}$ are real and nonzero. The eigenvalues then come in pairs of equal magnitude and opposite sign, as claimed previously. It can furthermore be shown\cite{Blaizot} that one can normalize the $q^2$ vectors $\Vv_{\kv}^\zeta$ so that
\beq{Vnormalization}
{\Vv_{\kv}^\zeta}^\dagger \etam \Vv_{\kv}^{\zeta'} = \delta_{\zeta\zeta'}\punc{,}
\eeq
and hence ${\Wv_{\kv}^\zeta}^\dagger \etam \Wv_{\kv}^{\zeta'} = -\delta_{\zeta\zeta'}$ and ${\Wv_{\kv}^\zeta}^\dagger \etam \Vv_{\kv}^{\zeta'} = 0$.

These orthonormality relations immediately lead to the results used in Section~\ref{SecBogvQuasiparticles}. First, they imply that the operators $d_{\kv\zeta} = {\Vv_{\kv}^\zeta}^\dagger \etam \alphav_{\kv} = -\alphav^\dagger_{-\kv}\etam\Wv_{-\kv}^\zeta$ obey the canonical commutation relations,
\beq{dcommutators}
[d\nd_{\kv\zeta},d^\dagger_{\kv'\zeta'}] = (2\pi)^2\delta^2(\kv-\kv')\delta_{\zeta\zeta'}\punc{.}
\eeq
Second, they lead to the inverse expression
\beq{alphafromd}
\alphav_{\kv} = \sum_{\zeta} \left(
\Vv_{\kv}^\zeta d\nd_{\kv \zeta} + \Wv_{\kv}^{\zeta} d^\dagger_{-\kv\, \zeta}
\right)\punc{,}
\eeq
giving
\beq{Hamdk}
\alphav^\dagger_{\kv} \Mm\nd_{\kv} \alphav\nd_{\kv} = \sum_\zeta \xi_{\kv\zeta} \left( d^\dagger_{\kv\zeta} d\nd_{\kv\zeta} + d\nd_{\kv\zeta} d^\dagger_{\kv\zeta}\right)\punc{,}
\eeq
from which \refeq{H2Bogv} follows.

\subsection{Near zero momentum}
\label{SecNearZeroMomentum}

The invariance of the mean-field energy $h_0$ under changes of phase of $A_{\ell\gamma}$ leads to a vanishing eigenvalue of the matrix $\Mmzero$, which is by assumption the only zero eigenvalue (generically the case when no further continuous symmetries are broken). The corresponding eigenvector is given by
\beq{DefineP}
\Pv_{\ell\gamma} =\frac{1}{\sqrt{2|\Av|^2}}
\begin{pmatrix}
\ii A\ns_{\ell\gamma}\\
-\ii A^*_{\ell\gamma}
\end{pmatrix}\punc{,}
\eeq
and is obviously also an eigenvector of $\etam \Mmzero$ with zero eigenvalue. We choose the normalization and phase of $\Pv$ so that $\Pv^\dagger\Pv = 1$ and $\Pv = \pim\Pv = \gammam\Pv^*$.

It is convenient to define the vector $\Qv$ satisfying
\begin{align}
\label{Qdefinition1}
\etam\Mm_{\zerov}\Qv &= -\ii\mut\Pv\\
\label{Qdefinition2}
\Qv^\dagger \etam \Pv &= \ii\\
\label{Qdefinition3}
\Qv^\dagger \Pv &= 0\punc{,}
\end{align}
where $\mut$ is a constant with dimensions of energy. The vector $\Qv$ is specified uniquely by these three equations, because $\Mmzero$ has only one zero eigenvalue, and so its inverse can be defined in the subspace orthogonal to $\Pv$. One can therefore write $\Qv = -\ii \mut \Mmzero^{-1} \etam\Pv$, because $\etam\Pv$ is orthogonal to $\Pv$, and so is $\Qv$ by \refeq{Qdefinition3}. \refeq{Qdefinition2} simply fixes the normalization of $\Qv$, and thus the constant $\mut$, which can be shown\cite{Blaizot} to be positive and given by
\beq{mut}
\mut^{-1} = \Pv^\dagger \etam \Mmzero^{-1} \etam \Pv\punc{.}
\eeq

The vectors $\Pv$ and $\Qv$ are `conjugate' in the sense that they obey \refeq{Qdefinition2} along with
\beq{PQorthogonality}
\Pv^\dagger \etam \Pv = \Qv^\dagger \etam \Qv = 0\punc{,}
\eeq
so that operators constructed using the vectors $\Pv$ and $\Qv$ obey the commutation relations of momentum and position. These operators, however, apply only precisely at the zero-measure point $\kv = \zerov$, and so are not directly relevant for the properties in the thermodynamic limit.

Together with the eigenvectors $\Vv^\zeta_{\zerov}$ and $\Wv^\zeta_{\zerov}$ corresponding to nonzero eigenvalues, $\Pv$ and $\Qv$ span the $2q^2$-dimensional vector space, and can therefore be used as a basis for a perturbative description of the region near $\kv = \zerov$. For small $\kv$ (along $\xhv$, say), the matrix $\Mm_{\kv}$ can be expanded as
\beq{Mmexpand}
\Mm_{k_x \xhv} \simeq \Mmzero + k_x \Mm^{(1)} + k_x^2 \Mm^{(2)}\punc{,}
\eeq
and $\Mm^{(1,2)}$ can be treated as perturbations. The coefficient matrices can themselves be calculated using perturbation theory for the wavefunctions $\psi_{\gamma n}(\kv)$. The resulting expressions are analytic functions of the momentum $\kv$, implying the important symmetry property that $\Pv^\dagger \Mm^{(1)} \Pv = \Qv^\dagger \Mm^{(1)} \Qv = 0$.

In the following, we restrict to cases where the condensate is sufficiently symmetric that the phonon velocity is isotropic. This requires a symmetry of $\Av$ under either $\R$ or $\Ixy$ (see Appendix~\ref{AppSymmetries}), and as seen in Table~\ref{TabConfigurations}, is always the case for $q \le 5$. (The extension to general symmetry is straightforward.)

Because of the nonhermitian nature of the matrix $\etam \Mm_{\kv}$, standard Rayleigh-Schr\"odinger perturbation theory cannot be applied directly to this problem, and the resulting expressions for the eigenvalues and -vectors are not analytic in $\kv$. Instead, the smallest eigenvalue $\xi_{\kv1}$ is linear in $|\kv|$,
\beq{xismallk}
\xi_{\kv1} = c |\kv| + \mathcal{O}(|\kv|^2)\punc{,}
\eeq
with phonon velocity $c$ given by
\beq{PhononVelocity}
\frac{c^2}{\mut} = \Pv^\dagger \Mm^{(2)} \Pv - 2\sum_\zeta \frac{|\Pv^\dagger \Mm^{(1)} \Vv_{\zerov}^\zeta|^2}{\xi_{\zerov \zeta}}\punc{.}
\eeq

As described in Section~\ref{SecBogvNonzero}, away from $\kv = \zerov$ all eigenvectors have nonzero eigenvalues and can be normalized so that ${\Vv_{\kv}^\zeta}^\dagger\etam\Vv_{\kv}^\zeta = -{\Wv_{\kv}^\zeta}^\dagger\etam\Wv_{\kv}^\zeta = 1$. As $\kv$ approaches $\zerov$, the vectors $\Vv_{\kv}^1$ and $\Wv_{\kv}^1$ corresponding to the smallest eigenvalue both approach $\Pv$, the unique zero-eigenvector of $\Mmzero$. Since this eigenvector satisfies $\Pv^\dagger\etam\Pv = 0$, the normalization of both $\Vv_{\kv}^1$ and $\Wv_{\kv}^1$ must diverge as $\kv\rightarrow \zerov$. To leading order, one finds
\beq{Vsmallk}
\Vv_{\kv}^1 = v_0 |\kv|^{-1/2}\Pv + \mathcal{O}(|\kv|^{1/2})\punc{,}
\eeq
with coefficient $v_0 = \sqrt{\mut/(2c)}$. The omitted higher-order terms maintain the appropriate normalization: $\Pv^\dagger\etam\Vv_{\kv}^1 = v_0 |\kv|^{1/2} c/\mut + \mathcal{O}(|\kv|^{3/2})$.

\section{Symmetries and degeneracies}
\label{AppSymmetries}

In Section~\ref{SecMSG}, the magnetic symmetry group (MSG) was introduced, and the operators corresponding to various elementary operations were defined. In this Appendix, we will present in detail the consequences of these symmetries for the condensate configurations and the quasiparticle spectrum.

We denote the full group of symmetries of the Hamiltonian $\Ham$ as $\MSG$, which includes the translations, rotations, and reflections considered in Section~\ref{SecMSG}. As noted in Section~\ref{SecMFTheory}, the mean-field condensate configuration breaks a subset of $\MSG$; the subgroup of symmetries that are preserved will be denoted $\MSGr$. Examples are given in Table~\ref{TabConfigurations}, which lists the symmetries preserved in the configurations illustrated in Figure~\ref{FigCurrents}. We will discuss the various properties of the spectrum, including symmetries and degeneracies, that result from $\MSGr$.

A general spatial transformation, such as an element of $\MSG$, is represented by the (anti)unitary operator $\S$, under which $a_{\kv \ell \gamma}$ transforms as
\beq{Sak}
\S a_{\kv \ell \gamma} = \sum_{\ell',\gamma'} s_{\ell \gamma,\ell'\gamma'}(\kv) a_{(\SS \kv)\ell'\gamma'} \S\punc{.}
\eeq
It should be recalled that reflection operators must be combined with time reversal to give symmetries of the Hamiltonian, and are hence represented by antiunitary operators. The coefficients in \refeq{Sak}, which can be written as a $q^2 \times q^2$ matrix $\sm_{\kv}$, are diagonal in $\gamma$ (with possible exceptions at points where two bands touch). Preservation of the commutation relations requires that $\sm_{\kv}$ be a unitary matrix, including for antiunitary $\S$.

Under the general operation $\S$, the condensate configuration $\Av$ is mapped to $\sm^{-1}(\zerov) \Av^{(*)}$, with complex conjugation if $\S$ is antiunitary. (The inverse matrix arises from considering transformations of states versus operators.) The mean-field energy $h_0$, defined in \refeq{MFT}, is therefore symmetric under the same transformations of $A_{\ell\gamma}$ as the full Hamiltonian is under transformations of $a_{\zerov \ell \gamma}$. A given configuration that minimizes $h_0$ will in general have lower symmetry, however, being invariant only under (a group of mappings isomorphic to) $\MSGr$. As argued in Section~\ref{SecMFTheory}, the superfluid therefore breaks spatial symmetries as well as the $\mathrm{U}(1)$ phase symmetry.

As usual, the broken symmetries, comprising the subset $\MSG \setminus \MSGr$, imply the existence of multiple degenerate configurations. For example, $h_0$ is invariant under $A_{\ell\gamma} \rightarrow A_{[\ell-1]_q\gamma}$ and under $A_{\ell\gamma} \rightarrow \omega^{-\ell} A_{\ell\gamma}$, corresponding to $\Tx$ and $\Ty$ respectively [see \refeqand{Tyakl}{Txakl}]. These two mappings do not commute, and so there is no configuration $\Av$ that preserves both symmetries. This leads to the conclusion that the number of distinct configurations that minimize $h_0$ is always a multiple of $q$ (by the same argument that implies the degeneracy of the single-particle states with different $\ell$).

The consequences of the symmetries $\MSGr$ for the quasiparticle spectrum can be determined by considering the transformations of the quadratic matrix $\Mm_{\kv}$. The single-particle contribution to $\Mm_{\kv}$, given by the first term in \refeq{QuadraticMatrix}, commutes with any operation $\S$ in the full symmetry group $\MSG$, while the second term results from interactions with the condensate, and so is symmetric only under the elements of $\MSGr$.

These preserved symmetries imply constraints on the matrix $\Mm_{\kv}$. In particular, defining $\sm$ as in \refeq{Sak} and taking $\S$ to be (anti)unitary, one can show that if $\sm_{\zerov}^{-1} \Av^{(*)} = \sigma \Av$, then $\Sigmam\nd_{\kv}\Mm_{\kv}^{(*)} = \Mm\nd_{\SS\kv} \Sigmam\nd_{\kv}$, where
\beq{Sigmam}
\Sigmam_{\kv} =
\begin{pmatrix}
\sigma^* \sm_{\kv}^\dagger & 0\\
0 & \sigma \sm_{-\kv}^\tr
\end{pmatrix}\punc{.}
\eeq
In the presence of a condensate that is symmetric under a transformation $\S$, the quasiparticle energies are therefore equal at $\kv$ and $\SS \kv$ (including in the case where $\S$ is antiunitary). For antiunitary operators, it is convenient to use the notation $\check{\Sigmam}$ for the operation of complex conjugation followed by multiplication by the matrix $\Sigmam$.

Applied to momenta near $\kv = \zerov$, this leads to constraints on the phonon speed $c$, calculated in Section~\ref{SecNearZeroMomentum}. In all four cases listed in Table~\ref{TabConfigurations}, the symmetry is sufficient to have isotropic phonon speed, as assumed in \refeq{xismallk}. For $\alpha = \frac{1}{2}$ and $\frac{1}{3}$, the relevant symmetry is the (antiunitary) reflection $\Ixy$, while for $\alpha = \frac{1}{4}$ and $\frac{1}{5}$, it is the rotation $\R$. (For $\alpha = \frac{1}{2}$, there is also symmetry under rotation by $\frac{\pi}{2}$ about a plaquette center, $\Tx\R$.)

In all cases, the spectrum is of course symmetric only under, at most, the fourfold rotation symmetry of the square lattice. The symmetry under {\em continuous} rotations of $\kv$ in \refeq{xismallk} is a simple example of an emergent low-energy symmetry.

In many cases (for example $\alpha = \frac{1}{2}$, $\frac{1}{3}$, and $\frac{1}{5}$; see Figure~\ref{FigCurrents} and Table~\ref{TabConfigurations}), the condensate configuration preserves a nontrivial translation symmetry. In particular, suppose translation $\TR$ by a displacement $\Rv$ is unbroken, where $\Rv$ is not a lattice vector of the enlarged unit cell. (Because $\Tx$ and $\Ty$ do not commute, one must specify the path to define $\TR$ precisely.) Since translation does not change $\kv$, one can define an operator using \refeq{Sigmam} that commutes with $\Mm_{\kv}$. For each $\kv$, the modes $\zeta$ can therefore be labeled according to their eigenvalue under $\Sigmam_{\kv}$. Modes with different eigenvalues are allowed to have (unavoided) crossings, as visible for example in Figure~\ref{FigDispersions13}.

\subsection{High-symmetry points}

Points in momentum space separated by the reciprocal lattice vectors $2\pi\xhv$ and $2\pi\yhv$ are physically equivalent, so the full Brillouin zone $\BL$ has the topology of a torus. The same applies, with some modifications, to the doubly reduced Brillouin zone $\BN$.

With momentum shift operators defined by $\dsK_x \kv = [\kv + \frac{2\pi}{q}\xhv]_{\BL}$ and $\dsK_y \kv = [\kv + \frac{2\pi}{q}\yhv]_{\BL}$, the matrix $H_{nn'}(\kv)$, defined in \refeq{Hmatrix}, obeys
\begin{align}
\label{HnnX}
H_{nn'}(\dsK_x\kv) &= H_{[n+\pb]_q[n'+\pb]_q}(\kv)\\
\label{HnnY}
H_{nn'}(\dsK_y\kv) &= \omega^{-\pb (n - n')}H_{nn'}(\kv)\punc{,}
\end{align}
so one can extend the definition of $\psi_{\gamma n}(\kv)$ beyond $\BN$ by
\begin{align}
\label{psiX}
\psi_{\gamma n}(\dsK_x\kv) &= \psi_{\gamma [n+\pb]_q}(\kv) \ee^{\ii \theta_\gamma^x(\kv)}\\
\label{psiY}
\psi_{\gamma n}(\dsK_y\kv) &= \omega^{-\pb n}\psi_{\gamma n}(\kv) \ee^{\ii \theta_\gamma^y(\kv)}\punc{.}
\end{align}

The phases $\theta_\gamma^{x,y}(\kv)$, corresponding to the flux threaded through the holes of the torus, are arbitrary apart from constraints due to symmetry. These can be found by considering the commutation relations of the operators $\dsK_{x,y}$ with each other and with the symmetry operators, and by requiring that $\psi_{\gamma n}(\kv)$ be continuous (apart from possibly at degeneracy points). The definitions in \refeqand{psiX}{psiY} are particularly useful on the boundary of $\BN$, where the number of constraints on $\theta_\gamma^{x,y}(\kv)$ is larger, and especially at the high-symmetry points at the corner (X) and edge-center (M) of $\BN$.

The relations between the eigenvectors at $\kv$ and $\dsK_\mu \kv$ lead to corresponding relations for $\Mm_{\kv}$, given by $\Km^\mu_{\kv} \Mm_{\kv} = \Mm_{\dsK_\mu \kv} \Km^\mu_{\kv}$, where
\begin{align}
\Km^x_{\ell\gamma,\ell'\gamma'}(\kv) &= \delta_{\ell\ell'}\delta_{\gamma\gamma'} \omega^{-\pb\ell}
\begin{pmatrix}
\ee^{-\ii\theta^x_\gamma(\kv)}&0\\
0&\ee^{\ii\theta^x_\gamma(\kv)}
\end{pmatrix}
\label{Kmx}\\
\Km^y_{\ell\gamma,\ell'\gamma'}(\kv) &= \delta_{\gamma\gamma'} 
\begin{pmatrix}
\ee^{-\ii\theta^y_\gamma(\kv)}\delta_{\ell',[\ell+\pb]_q}&0\\
0&\ee^{\ii\theta^y_\gamma(\kv)}\delta_{\ell',[\ell-\pb]_q}
\end{pmatrix}
\punc{.}
\end{align}
These immediately imply that the mode energies are equal at points on opposite sides of $\BN$.

\subsection{Kramers degeneracy}
\label{SecKramers}

For $\alpha = \frac{1}{2}$, there is a twofold degeneracy at every point along the line from M to X, as can be seen in Figure~\ref{FigDispersions12}, and by symmetry at every point on the edge of $\BN$. (The same degeneracy also occurs for $\alpha = \frac{1}{4}$.)

This is in fact a Kramers degeneracy, and is a consequence of the symmetry under the glide reflection $\Tx\Iy$ (for $\alpha = \frac{1}{2}$; the corresponding symmetry for $\alpha = \frac{1}{4}$ is $\Ty^2\Tx^2\Iy$). Its action in momentum space is to map $\kv$ to $\IIx \kv$, so that a point $\kv = \frac{\pi}{q}\xhv + k_y \yhv$, on the line from M to X, is mapped to $-\frac{\pi}{q}\xhv + k_y \yhv$, on the opposite side of $\BN$. The quadratic matrix $\Mm_{\kv}$ at such a point therefore obeys $[\check{\mathbf{\Gamma}}_{\kv}, \etam\Mm_{\kv}] = \zerov$, where
\beq{MXsymmetry}
\mathbf{\Gamma}_{\kv} = \Km^x_{\IIx\kv} \Sigmam_{\kv}^{\Tx \Iy}\punc{,}
\eeq
with $\Sigmam_{\kv}^{\Tx \Iy}$ defined in \refeq{Sigmam} and $\Km^x_{\kv}$ in \refeq{Kmx}.

The antiunitary operator $\check{\mathbf{\Gamma}}_{\kv}$ has the property
\beq{SigmaSquared}
\check{\mathbf{\Gamma}}_{\kv}^2 = \mathbf{\Gamma}_{\kv}\ns 
\mathbf{\Gamma}_{\kv}^* = -\boldsymbol{1}\punc{,}
\eeq
which implies, by Kramers' theorem, that all eigenvalues of $\etam\Mm_{\kv}$ are twofold degenerate. Note that, in the case $\alpha = \frac{1}{4}$, this Kramers degeneracy exists despite the explicit breaking of time-reversal symmetry by the applied magnetic field. (For $\alpha = \frac{1}{2}$, $\omega = -1$ is real, and the Hamiltonian preserves time-reversal symmetry. The condensate configuration nonetheless breaks this symmetry, as shown in Figure~\ref{FigCurrents}.)

\section{Analytics for $\alpha = \frac{1}{2}$}
\label{AppAnalytics}

The simplest nontrivial case is $\alpha = \frac{1}{2}$, and it is then possible to perform many of the calculations analytically. In this case, the matrix $\Hm(\kv)$ defined in Section~\ref{SecMomentumSpace} is given by
\beq{Hmq2}
\Hm(k_x \xhv + k_y \yhv) = -2t
\begin{pmatrix}
\cos k_x & \cos k_y\\
\cos k_y & -\cos k_x
\end{pmatrix}\punc{,}
\eeq
with eigenvalues
\beq{epsilon2}
\epsilon_{\kv} = \pm 2t\sqrt{\cos^2 k_x + \cos^2 k_y}
\punc{.}
\eeq
Note that the spectrum has a Dirac cone at the corner of $\BN$, where $|k_x| = |k_y| = \frac{\pi}{2}$. At $\kv = \zerov$, the eigenvectors are, for bands $\gamma \in \{1,2\}$,
\beq{Hmq2evecs}
\boldsymbol{\psi}_\gamma(\zerov) = \frac{1}{\sqrt{4+2(-1)^\gamma\sqrt{2}}}
\begin{pmatrix}
1+(-1)^\gamma\sqrt{2}\\
1
\end{pmatrix}\punc{.}
\eeq

The mean-field condensate configuration can be determined by calculating $h_0$, given in \refeq{MFT}, and minimizing with respect to $A_{\ell\gamma}$ at fixed average density. In this case, however, the appropriate configuration is more easily found by inspection. With the choice $A_{\ell,1} = \ii^\ell\sqrt{\rho/2}$ and $A_{\ell,2} = 0$, direct calculation shows that the real-space wavefunction, given by \refeq{RealSpaceWavefunction}, has uniform magnitude,
\beq{RSWf2}
\langle b_j \rangle = \sqrt{\rho} \exp \left\{\ii (-1)^{y_j}[(-1)^{x_j}\frac{\pi}{4} - \frac{5\pi}{8}] \right\}\punc{.}
\eeq
This configuration, in which the condensate is restricted to the lower band, therefore has uniform density of $\rho$ particles per lattice site. Calculation of the currents using \refeq{CurrentOperator} gives configurations as illustrated in Figure~\ref{FigCurrents}, with the current on each link having magnitude $|\langle \J \rangle| = \sqrt{2}t \rho$. Replacing $A_{\ell\gamma}$ by its complex conjugate gives the equivalent configuration with currents reversed on each link.

Since these configurations have uniform density, they simultaneously minimize both terms in \refeq{MFT} globally, and are therefore global minima of $h_0$, at fixed density $\rho$. (They minimize the kinetic energy because they contain contributions only from the degenerate minima of the lowest band, and they minimize the potential energy because they have uniform density.) The case $\alpha = \frac{1}{2}$ is unique in this regard, with the condensate configuration minimizing both terms simultaneously and so insensitive to the value of the interaction strength $U$. For $q > 2$, the density modulations can be reduced by including higher bands in the condensate configuration, and the extent to which they contribute is determined by $U / t$.

These two configurations are in fact the only minima (up to a redundant overall phase rotation), and provide an example of the general result that there is always a discrete set of degenerate minima, whose number is a multiple of $q$. Either of the translation operators $\Tx$ and $\Ty$ relates one of the two configurations to the other, up to an overall phase (using the transformation of $A_{\ell\gamma}$ specified in Appendix~\ref{AppSymmetries}).

Both configurations are symmetric under $\Ty \Tx$, as noted in Table~\ref{TabConfigurations}, with the first obeying
\beq{EqTransSymm2}
\sum_{\ell\ell'} \left(\mathbf{s}^{\Ty\Tx}\right)^{-1}_{\ell\ell'} A_{\ell',1} = \ii A_{\ell,1}\punc{,}
\eeq
where $\mathbf{s}^{\Ty\Tx} = \begin{pmatrix}0&1\\-1&0\end{pmatrix}$ is the matrix defined by \refeq{Sak} for the transformation $\Ty\Tx$ (at $\kv = \zerov$ and restricted to $\gamma = 1$). One can therefore construct the matrix $\Sigmam^{\Ty\Tx}$ according to \refeq{Sigmam}; it has eigenvalues $\pm 1$, allowing the quasiparticle modes to be labeled as even or odd under $\Ty\Tx$.

\subsection{Quasiparticle dispersion}

To find the dispersion, one must construct the $8\times 8$ matrix $\Mm_{\kv}$ given in \refeq{QuadraticMatrix}. Rather than using the single-particle basis labeled by $\kv$, $\ell$, and $\gamma$, it is somewhat easier to find analytic results by starting in the basis of $\kv$, $\ell$, and $n$ as in \refeq{Hamtk2}. While the first term in \refeq{QuadraticMatrix} is not diagonal in this basis, the second has a considerably simpler expression, as a result of the simple form of the on-site interaction in momentum space.

After transforming to the basis of eigenvectors of $\Sigmam^{\Ty\Tx}$, the matrix $\Mm_{\kv}$ splits into two $4 \times 4$ blocks,
\begin{widetext}
\beq{Mm2}
\Mm_{\kv}^{(\pm)} =
\begin{pmatrix}
 U \rho +2 t (\sqrt{2} + \cos k_x) & 2 t \cos k_y & \pm U \rho/\sqrt{2} & U \rho/\sqrt{2} \\
 2 t \cos k_y & U \rho + 2 t (\sqrt{2} - \cos k_x)  & U \rho/\sqrt{2} & \mp U \rho/\sqrt{2} \\
 \pm U \rho/\sqrt{2} & U \rho/\sqrt{2} & U \rho +2 t (\sqrt{2} + \cos k_x) & 2 t \cos k_y \\
U \rho/\sqrt{2} & \mp U \rho/\sqrt{2} & 2 t \cos k_y & U \rho + 2 t (\sqrt{2} - \cos k_x) 
\end{pmatrix}\punc{,}
\eeq
corresponding to the eigenvalues $\pm1$. It is then straightforward to calculate the eigenvalues of $\etam \Mm_{\kv}$ as discussed in Section~\ref{SecInversionSymmetry}. They come in pairs of opposite sign, and so their squares are given by the roots of a quadratic equation.

The quasiparticle energies $\xi_{\kv}$ are finally given by
\beq{Eqxi2}
\xi_{\kv}^2 = 8 t^2+4 \sqrt{2} t U \rho + \epsilon_{\kv}^2 \pm \sqrt{(32 t^2+16 \sqrt{2} t U \rho + 2U^2 \rho ^2)\epsilon_{\kv}^2 \pm 16 t^2 U^2 \rho ^2 \cos k_x \cos k_y}\punc{,}
\eeq
\end{widetext}
where the two choices of $\pm$ are independent, giving the $q^2 = 4$ modes of the interacting dispersion. Taking the first sign as $-$ and the second as $+$ gives the Goldstone mode, which has $\xi_{\kv} = |\kv|\sqrt{\sqrt{2}\rho U t} + \mathcal{O}(|\kv|^3)$ near $\kv = \zerov$.

Using \refeq{Eqxi2}, one can confirm the twofold degeneracy along the line from M to X established in Section~\ref{SecKramers}. For these points, $\cos k_x = 0$ and the second choice of $\pm$ is redundant.

\subsection{Amplitude-phase description of gapless mode}
\label{SecAmplitudePhase}

The small-$|\kv|$ dispersion of the Goldstone mode can be derived by more elegant means if one restricts to long-wavelength fluctuations of the condensate configuration. Gradual variations of the real-space wavefunction can be parametrized by writing
\beq{bAmplitudePhase}
b_j = \langle b_j \rangle \ee^{\ii \vartheta_j} \sqrt{1 + \frac{\varrho_j}{\rho}}\punc{,}
\eeq
where $\vartheta_j$ and $\varrho_j$ describe deviations in the phase and amplitude respectively, and have canonical commutation relations $[\vartheta_i,\varrho_j] = \ii\delta_{ij}$.

We now rewrite the Hamiltonian $\Ham$ in terms of these new degrees of freedom. Assuming $\varrho_j \ll \rho$ and that both $\varrho_j$ and $\vartheta_j$ vary only over distances large compared to the lattice scale, one can expand to give
\beq{HamAmplitudePhase}
\Ham = h_0 + \frac{t \rho}{\sqrt{2}} \sum_{\langle i j \rangle} (\vartheta_i - \vartheta_j)^2 + \frac{U}{2}\sum_j \varrho_j^2 + \cdots\punc{.}
\eeq
Note that the frustration in $\Ham_t$ implies that the kinetic energy of each link is not separately minimized in the mean-field configuration. Each link $i\link j$ therefore contributes a term linear in $\vartheta_i - \vartheta_j$, but their sum vanishes, since the mean-field configuration is a minimum of the total kinetic energy.

Writing this equation in terms of the Fourier components of $\vartheta_j$ and $\varrho_j$, we obtain
\beq{HamAmplitudePhaseFT}
\Ham = h_0 + \int\!\dkv \left(\frac{t\rho}{\sqrt{2}}|\kv|^2 |\vartheta_{\kv}|^2 + \frac{U}{2}|\varrho_{\kv}|^2\right) + \cdots\punc{,}
\eeq
where the integral is restricted to small $|\kv|$ by the assumption of slowly varying fluctuations. This takes the form of a harmonic oscillator for each momentum, so the dispersion is $\xi_{\kv} = 2\sqrt{\frac{t\rho}{\sqrt{2}}|\kv|^2 \times \frac{U}{2}}$, in agreement with the result given in the previous section.


\begin{thebibliography}{99}
\bibitem{Footnote1} One can easily check that this gives the correct result for a classical particle that starts at position $\rv$ with zero kinetic energy. It has zero dynamical momentum, and hence canonical momentum $\scAv(\rv)$. When the Raman beams are suddenly switched off, its canonical momentum is preserved, so its final momentum (dynamical or canonical) is $\scAv(\rv)$. This momentum can be viewed as the result of an induced electric field $-\dot{\scAv}$ acting during the period that the gauge potential is switched off.\cite{Lin,LinNature}
\end{thebibliography}
\end{document}